\documentclass[aps, prd, amsmath, floats, floatfix, superscriptaddress, nofootinbib, showpacs]{revtex4-2}

\usepackage{aas_macros}
\usepackage{amssymb}
\usepackage{amsmath}
\usepackage{verbatim}
\usepackage{mathrsfs}
\usepackage{amsfonts}
\usepackage{latexsym}
\usepackage{epsfig}
\usepackage{epstopdf}
\usepackage{float}
\usepackage[usenames, dvipsnames]{color}
\usepackage{mathtools}
\usepackage{cases}
\usepackage{hyperref}
\usepackage{soul}
\usepackage{graphicx}
\usepackage{academicons}
\usepackage{svg}
\usepackage{subcaption}
\usepackage{graphicx}
\usepackage{caption}
\usepackage{ulem}

\newcommand{\dvol}{\mbox{dvol}}

\newcommand{\ve}[1]{{\underline #1}}

\begin{document}

\title{Entropy covector field and macroscopic observables for rotating and non-rotating relativistic kinetic gases around a Schwarzschild black hole}

\author{Carlos Gabarrete}
\email{carlos.gabarrete@unah.edu.hn}
\affiliation{Departamento de Gravitaci\'on, Altas Energ\'ias y Radiaciones, Escuela de F\'isica, Facultad de Ciencias, Universidad Nacional Aut\'onoma de Honduras, Edificio E1, Ciudad Universitaria, Tegucigalpa, Francisco Moraz\'an, Honduras}

\author{Daniela Montoya}
\email{dmmontoyaa@unah.hn}
\affiliation{Departamento de Gravitaci\'on, Altas Energ\'ias y Radiaciones, Escuela de F\'isica, Facultad de Ciencias, Universidad Nacional Aut\'onoma de Honduras, Edificio E1, Ciudad Universitaria, Tegucigalpa, Francisco Moraz\'an, Honduras}

\author{Roger Raudales}
\email{rraudales@upnfm.edu.hn}
\affiliation{Departamento de Ciencias Naturales, Facultad de Ciencias Básicas, Universidad Pedagógica Nacional Francisco Morazán, Edificio 3, Col. El Dorado, Tegucigalpa, Francisco Morazán, Honduras}

\date{\today}

\begin{abstract}
\centerline{\textsc{Abstract}} 
In this article, we derive the components of the entropy covector field for a relativistic kinetic gas composed of collisionless, spinless, massive, and uncharged particles following bound orbits in a curved spacetime background. By assuming a dependence on the inclination angle of the particle orbits, we consider two distinct models that describe a rotating and a non-rotating relativistic kinetic gas around a Schwarzschild black hole. We analyze the behavior of key macroscopic observables (including the anisotropy parameter and the kinetic temperature) which are constructed from the particle density, energy density, and principal pressures. We aim to characterize and compare the morphology of the resulting configurations, thereby extending and complementing a previous work. The results reveal significant differences between the rotating and non-rotating cases, particularly in the asymptotic behavior of the anisotropy parameter, kinetic temperature, and average pressure, highlighting the role of angular momentum in shaping the macroscopic properties of collisionless gases in strong gravitational fields.
\end{abstract}

\pacs{04.20.-q, 97.60.Lf, 05.20.Dd, 05.70.-a}

\maketitle

\section{Introduction}

The kinetic theory is based on the principles of statistical mechanics and explains the macroscopic properties of the gas (such as pressure, temperature, entropy, particle density, etc.) from microscopic laws of physics. This theory can be treated through one of the key concepts, namely, the characterization of the state of the gas through the one-particle distribution function (DF)~\cite{CercignaniKremer-Book}. The DF is a time-dependent function defined on the one-particle phase space of the theory, whose time evolution is determined by Boltzmann's equation~\cite{oStZ13,oStZ14a,oStZ14b}. The DF represents the averaged number of particles contained inside the volume element $d^3x\,d^3p$ as $dN = f d^3x d^3p$ at the nonrelativistic level.

This article focuses on the study of the behavior of a relativistic kinetic gas in a strong gravitational field. In particular, we are interested in kinetic gas clouds that propagate around black holes, and we explore different scenarios describing such configurations. We consider the case in which collisions between the gas particles, as well as effects of the self-gravity of the gas and the electromagnetic field, can be neglected as a starting point. This leads to the problem of solving the relativistic collisionless Boltzmann equation for the one-particle distribution function on a fixed spacetime background. Restricting ourselves to a spherically symmetric Schwarzschild black hole and choosing a DF depending solely on integrals of motion, the collisionless Boltzmann equation is solved automatically. From this one-particle DF of a simple gas, one can construct physical observables, through momentum averaging, that contain information about the particle, energy, and entropy densities; mean particle four-velocity; heat flow; pressure tensor; and the kinetic temperature.

For the development of this work, we assume a particular ansatz for the one-particle DF that describes a collisionless kinetic gas around a non-rotating black hole, motivated by the inclination angle of the particle's orbits~\cite{BinneyTremaine-Book,cGoS2022,cGrR2025} and the generalized polytropic ansatz~\cite{eAhAaL16,eAhAaL19}. The dependence on the inclination angle allows us to consider two configurations of a relativistic kinetic gas: one with total angular momentum (rotating model) and another with zero angular momentum (non-rotating model). Since we are working with a collisionless gas, it is not possible to establish thermal equilibrium in any sense. Consequently, the distribution function, which is a solution of the Liouville vector field, can be written as a function that depends solely on the constants of motion.

The main purpose of this article is to study and characterize the morphology of a collisionless relativistic kinetic gas around a Schwarzschild black hole through the analysis of macroscopic observables derived from a distribution function that depends on the constants of motion (energy, total angular momentum, and azimuthal angular momentum), with special emphasis on the comparison between configurations with total angular momentum (rotating model) and without it (non-rotating model). As novel contributions, we present for the first time a complete analysis of the entropy flux covector field and the invariant entropy density for both models, revealing that the inclusion of angular momentum systematically reduces the entropy density throughout the entire radial domain, with a nontrivial entropic contrast that exhibits a minimum at intermediate radii and monotonic growth toward large distances. Furthermore, we show that the anisotropy parameter in the rotating model not only can cross zero and become positive (a feature absent in the non-rotating case) but also tends asymptotically to a constant positive value that depends exclusively on the parameter $s$, indicating a persistent residual anisotropy even at large distances from the black hole. The kinetic temperature also exhibits contrasting parametric sensitivities between both models: sensitive to $k$ but insensitive to $s$ in the non-rotating case, while in the rotating model it shows weak sensitivity to both parameters, with the dependence on $s$ becoming more noticeable in the asymptotic region. Finally, we provide a systematic comparison with a hydrodynamic description (the Polish doughnut model), showing that while the particle density and average pressure exhibit morphological agreement with the average pressure being remarkably robust across all models the temperature profiles display no correlation between the kinetic and fluid descriptions, a result that highlights the distinctive signatures of collisionless dynamics and the importance of choosing the appropriate model depending on the physical regime of interest.

In Section~\ref{Sec:Preliminaries} we summarize the principal properties of a Schwarzschild exterior spacetime, and from the DF, we recall the principal ideas of the collisionless Boltzmann equation for the construction of the spacetime observables such as the particle current density covector field, the energy-momentum-stress tensor field (whose expressions were derived in the paper~\cite{cGrR2025}), and the entropy flux covector field, analyzed in Section~\ref{Sec:InvariantEntropyDensity}. From these fields we can obtain the particle, energy, and entropy densities, as well as their respective velocities. We present an illustrative example of local thermodynamic equilibrium using a Maxwell-J\"uttner DF~\cite{fJ11a,fJ11b}. In addition, from the energy-momentum stress tensor, we can also obtain the pressure tensor, from which, using the principal pressures, we analyze the anisotropy parameter in Section IV, as well as the kinetic temperature in Section V, to complete the description of the morphology of a relativistic kinetic gas. Contrary to previous works such as~\cite{pMoA2021a,pMoA2021b,aGetal2021}, which works with unbounded trajectories, this one focuses on describing the steady state of a collisionless relativistic kinetic gas that is bounded by the gravitational potential of the black hole and follows geodesic trajectories around it.

Throughout the article, we present the results and the analysis of the morphology of the resulting configurations through the behavior of the macroscopic quantities explored. We compare the resulting behavior of spacetime observables between models with and without total angular momentum and develop gas configurations. In the paper~\cite{cGrR2025} we derive the expressions for the principal fields used here; however, we add the analysis of the entropy, kinetic temperature, and the anisotropy parameter to complete the description of the morphology of a relativistic kinetic gas. In Section~\ref{Sec:HydroModel}, we provide some qualitative comments comparing our kinetic results with those obtained from a hydrodynamic description, such as the Polish doughnut model~\cite{Rezzolla-Book}, highlighting the distinctive signatures of collisionless dynamics. Finally, in Section~\ref{Sec:Conclusions} we state our conclusions. We use the signature convention $(-,+,+,+)$ for the spacetime metric and geometrized units in which Newton's constant and the speed of light are one, i.e., $G_N = c = 1$.

\section{Preliminaries of spacetime observables}
\label{Sec:Preliminaries}

From the one-particle DF $f$ of a simple gas, one can construct physical observables which are the most important $C^\infty$-smooth tensor fields on the spacetime manifold $(\mathcal{M}, g)$. They are obtained by suitable fiber integrals over the momenta, and they are the particle current density covector field $J$, the energy-momentum-stress tensor field $T$, and the entropy flux covector field $S$. One can write these spacetime observables defined by the one-particle DF of a simple gas in terms of adapted local coordinates $(x^\mu, p_\mu)$ as follows:
\begin{eqnarray}
    \label{Eq:Jmu}
    J_\mu(x) &=& \int\limits_{P_x^+(m)} f(x,p) p_\mu \dvol_x(p), \\
    \label{Eq:Tmunu}
    T_{\mu\nu}(x) &=& \int\limits_{P_x^+(m)} f(x,p) p_\mu p_\nu \dvol_x(p), \\
    \label{Eq:Smu}
    S_\mu(x) &=& -k_B \int\limits_{P_x^+(m)} f(x,p)\log(A f(x,p)) p_\mu \dvol_x(p).
\end{eqnarray}
Here, $P_x^+(m)$ is the future mass hyperboloid defined by 
\begin{equation}
    P_x^+(m) := \left\{ p\in T_x^*(\mathcal{M}) \: : \: g_x^{-1}(p,p) = -m^2, \: \hbox{the vector dual to} \: p \: \hbox{is future-directed} \right\},
\end{equation}
and $\dvol_x(p)$ is  volume element given by 
\begin{equation}
    \label{Eq:VolumeForm}
    \dvol_x(p) := \frac{dp_{\hat{1}}\wedge dp_{\hat{2}}\wedge dp_{\hat{3}}}{\sqrt{m^2 + dp_{\hat{1}}^2 + dp_{\hat{2}}^2 + dp_{\hat{3}}^2}}
\end{equation}
in terms of an orthonormal basis of covector fields $\left\{\theta^{\hat{\alpha}}\right\}$ on $\mathcal{M}$. $k_B$ is the Boltzmann constant, and $A$ refers to an arbitrary positive constant of inverse units to those of the DF that ensures that the product $Af$ is dimensionless. Other observables can be obtained from the one-particle DF, including higher-order ones, but for the purposes of this article, the fields mentioned above are the only ones necessary. From these relevant fields~(\ref{Eq:Jmu}-\ref{Eq:Smu}), one can obtain information about the particle, energy, and entropy densities; mean particle four-velocity; heat flow; pressure tensor; and the kinetic temperature. 

From~(\ref{Eq:Jmu}) one can define the first two relevant quantities for our study, which are
\begin{eqnarray}
    \label{Eq:ParticleDensity}
    n(x) := \sqrt{-J^\mu (x) J_\mu (x)}, & &\qquad \hbox{(invariant particle density)}, \\
    \label{Eq:MeanVelocity}
    u_N^\mu(x) := \frac{1}{n(x)}J^\mu(x), & &\qquad \hbox{(mean particle velocity)}.
\end{eqnarray}
The invariant particle density $n$ can be understood as the number of particles per unit volume measured by an observer who is comoving with the mean particle flow $u_N$. Let $u_E$ be the timelike eigenvector of the energy-momentum-stress tensor~(\ref{Eq:Tmunu}) in which 
\begin{equation}
    T^{\mu}{}_{\nu}u_E^\nu = -\varepsilon u_E^\mu,
\end{equation}
this tensor can be diagonalized~\cite{Synge2-Book} and admits the following decomposition,
\begin{equation}
    T^{\mu}{}_{\nu}(x) = \varepsilon(x) (e_{\hat{0}})^\mu \otimes (e_{\hat{0}})_\nu + P_i(x) (e_{\hat{i}})^\mu \otimes (e_{\hat{i}})_\nu,
    \label{Eq:TmunuP}
\end{equation}
where $e_{\hat{0}} = u_E$ and $e_{\hat{i}}$ are perpendicular to $e_{\hat{0}}$. From this decomposition, $T$ one recognizes, on one hand, the energy density $\varepsilon$ defined as the negative of the eigenvalue corresponding to the timelike eigenvector of the energy-momentum-stress tensor, and on the other hand, the principal components of the pressure tensor $P_i$, defined as the eigenvalues belonging to the spacelike eigenvectors of the tensor field. 

From the entropy current density covector field~(\ref{Eq:Smu}) one can define 
\begin{eqnarray}
\label{Eq:EntropyDensity}
    S(x) := \sqrt{-S^\mu (x) S_\mu (x)}, & & \qquad \hbox{(invariant entropy density)}, \\
\label{Eq:EntropyVelocity}
    u_S^\mu(x) := \frac{1}{S(x)}S^\mu(x), & & \qquad \hbox{(entropy velocity)}.
\end{eqnarray}
By contracting the entropy current density~(\ref{Eq:Smu}) with the entropy velocity $u_S^\mu$ one obtains  
\begin{equation}
    \mathcal{S}(x) := -S_\mu(x) u_S^\mu(x).
\label{Eq:Sdensity}
\end{equation}
Analogous to the invariant particle density, the entropy density $\mathcal{S}$ is measured by a comoving observer with the entropy velocity $u_S$. This quantity is of special interest for thermodynamic considerations. We remark here that the three velocities $(u_N, u_E, u_S)$ defined by the fields~(\ref{Eq:Jmu}-\ref{Eq:Smu}) do not need to coincide in general. However, if local thermodynamical equilibrium is reached, these velocities are equal to each other. Taking the divergence of~(\ref{Eq:Jmu}) and~(\ref{Eq:Tmunu}), one can show that~\cite{rAcGoS2022}
\begin{eqnarray}
    \nabla^\mu J_\mu(x) &=& \int\limits_{P_x^+(m)} L[f](x,p) \dvol_x(p), \\
    \nabla^\mu T_{\mu\nu}(x) &=& \int\limits_{P_x^+(m)} L[f](x,p) p_\nu \dvol_x(p),
\end{eqnarray}
in which $L$ denotes the Liouville vector field. These divergences vanish if the Liouville equation $L[f]=0$ is satisfied; in this case
\begin{equation}
    \nabla^\mu J_\mu = 0, \qquad \nabla^\mu T_{\mu\nu} = 0, 
\end{equation}
are divergence free. Moreover, the divergence of the covector field~(\ref{Eq:Smu}) is
\begin{equation}
    \nabla^\mu S_\mu(x) = -k_B \int\limits_{P_x^+(m)} \left[1 + \log(Af(x,p))\right] L[f](x,p) \dvol_x(p),
\label{Eq:DivEntropy}
\end{equation}
which again vanishes if $f$ satisfies the Liouville equation. When collisions are included, one has
\begin{equation}
    \nabla^\mu S_\mu(x) \geq 0,
\end{equation}
as a consequence of Boltzmann's H-theorem~\cite{wI63}. The divergence free entropy flux implies that the entropy density is constant in time. When, $\nabla^\mu S_\mu = 0$ or its equivalent, $L[f]=0$ local thermodynamical equilibrium does not always describe the gas. 

As an illustrative example, we consider a gas configuration that is described by a Maxwell-J\"uttner distribution function~\cite{fJ11a,fJ11b} that determines a fluid in local equilibrium. For this purpose, one assumes the one-particle distribution function as follows
\begin{equation}
\label{Eq:Juttner}
    f(x,p) := \alpha(x) e^{\beta^\mu(x) p_\mu},
\end{equation}
where $\alpha$ is a positive function and $\beta$ is a future-directed timelike vector field on $\mathcal{M}$. The resulting macroscopic observables for a perfect fluid are 
\begin{eqnarray}
    h = \frac{e + P}{n}, &\quad& \hbox{(enthalpy per particle)}, \\
    P = \frac{m n}{z}, &\quad& \hbox{(hydrostatic pressure)}, \\
    u^\mu = \frac{m}{z} \beta^\mu, &\quad& \hbox{(mean particle velocity)},
\end{eqnarray}
where $e$ is the energy per particle and $z := m / (k_B T)$ is the ratio between the particle rest mass $m$ and the thermal energy (see, for example, ref.~\cite{rAcGoS2022}). For the distribution function~(\ref{Eq:Juttner}) the resulting entropy flux covector field is given by
\begin{eqnarray}
    S_\mu &=& -k_B\int\limits_{P_x^+(m)} f(x,p) \log(A f(x,p) ) p_\mu \dvol_x(p) \nonumber\\
    &=& -k_B \int\limits_{P_x^+(m)} \alpha e^{-\beta^\mu p_\mu} \left[ \log(A\alpha) + \beta^\nu p_\nu \right] p_\mu \dvol_x(p) \nonumber\\
    &=& -k_B \left[ \log(A\alpha) J_\mu + \beta^\nu T_{\mu\nu} \right],
\end{eqnarray}
hence, $S_\mu = s n u_\mu$, with an entropy per particle
\begin{equation}
    s = -k_B\left[ 1 + \log(A\alpha) \right] + \frac{h}{T},
\end{equation}
which satisfies the Gibbs relation (see for example~\cite{cGoS2023b} and references therein)
\begin{equation}
    ds = \frac{1}{T} d\left( \frac{\varepsilon}{n} \right) + \frac{P}{T} d\left( \frac{1}{n} \right).
\label{Eq:GibbsRelation}
\end{equation}

In the following section, we derive the explicit expression for the entropy covector field for a relativistic kinetic gas consisting of identical, uncharged, and spinless massive particles that follow bound orbits in the potential generated by the spherically symmetric curved spacetime background for a given model based on the inclination angle of the orbits and the polytrophic ansatz. 

The dependence on the inclination angle allows us to consider two configurations of a relativistic kinetic gas: one with total angular momentum (rotating model) and another with zero angular momentum (non-rotating model). We analyze and compare the results obtained for both models. Since we are working with a collisionless gas, it is not possible to establish thermal equilibrium in neither sense. Consequently, the distribution function, which is a solution of the Liouville vector field, can be written as a function that depends only on the constants of motion. In the following section~\ref{Sec:InvariantEntropyDensity}, we will adopt a distribution function model for this non-collisionless gas, which allows us to describe the behavior of the macroscopic observables of interest that will be explored later in sections~\ref{Sec:Anisotropy} and~\ref{Sec:KTemperature}.

\section{Profiles of the invariant entropy density}
\label{Sec:InvariantEntropyDensity}

Our system consists of a relativistic collisionless kinetic gas around a non-rotating black hole, in which the collisionless Boltzmann equation given by
\begin{equation}
    L[f] = 0,
\end{equation}
is satisfied by any one-particle DF depending only on integrals of motion. These conserved quantities associated with the spacetime manifold are the mass of the particles $m$, the energy $E$, the angular momentum of the gas particles $L:=|\ve{L}|$, and the azimuthal angular momentum $L_z$. Based on this, we assume an ansatz for the one-particle DF motivated by the generalized polytrophic ansatz for the dependency of the energy function and the inclination angle of the particle’s orbits~\cite{BinneyTremaine-Book,eAhAaL16,eAhAaL19,cGoS2022,cGrR2025}. DF dependency of the inclination angle allows us to infer a relativistic kinetic gas with (and without) total angular momentum, that is, a kinetic gas rotating (or not) around a Schwarzschild black hole. 

As mentioned above and using the models presented in the previous works~\cite{cGoS2022,cGoS2023b,cGrR2025}, we assume the one-particle DF as a product of an energy function times an inclination angle function
\begin{equation}
    \label{Eq:OneParticleDistrFunction}
    \mathcal{F}_m(E,L,L_z) := F_0(E) \times G(i).
\end{equation}
We introduce the following convenient dimensionless parametrization:
\begin{equation}
    \label{Eq:Scale}
    \xi := \frac{r}{M}, \quad 
    \lambda := \frac{L}{m M}, \quad 
    \lambda_z := \frac{L_z}{m M}, \quad
    \varepsilon := \frac{E}{m}, \quad
    \varepsilon_0 := \frac{E_0}{m}, \quad \hbox{and} \quad 
    U_\lambda(\xi) := \frac{V_{m,L}(r)}{m^2}, 
\end{equation}
where $V_{m,L}$ is the radial effective potential associated with the Schwarzschild spacetime, and it is defined by
\begin{equation}
    \label{Eq:SchEffectivePotential}
    V_{m,L}(r) = \left(1-\frac{2M}{r} \right) \left(m^2 + \frac{L^2}{r^2} \right), \qquad \hbox{and} \qquad
    U_{\lambda}(\xi) = N(\xi) \left(1 + \frac{\lambda^2}{\xi^2} \right),
\end{equation}
with the function $N$ given by $\displaystyle N(\xi) := 1 - \frac{2}{\xi} > 0$. The function depending on the energy function is motivated by the generalized polytrophic ansatz and is given by
\begin{equation}
    \label{Eq:F0}
    F_0(\varepsilon) = \alpha \left(1 - \frac{\varepsilon}{\varepsilon_0} \right)_+^{k-\frac{3}{2}},
\end{equation}
where $k > 1/2$, $\alpha > 0$, and $\varepsilon_0 > 0$ are constants. The parameter $\varepsilon_0 \leq 1$ is an energy cutoff that provides an upper bound for the energy. When $\varepsilon_0 < 1$ the configurations have finite extent (see~\cite{cGrR2025} for details). The notation $\mathcal{D}_+$ refers to the positive part of the quantity $\mathcal{D}$, that is, $\mathcal{D}_+ = \mathcal{D}$ if $\mathcal{D} > 0$ and $\mathcal{D}_+ = 0$ otherwise. 

For the function, $G \equiv G^{(\textrm{even},\textrm{rot})}_{i,i/2}$ depending on the inclination angle $i$ defined by $\displaystyle \cos i = L_z/L = \lambda_z/\lambda$, we assume two models, one that describes a non-rotating gas (even-$i$) and the other that describes a rotating gas (rot-$i/2$), defined as follows
\begin{eqnarray}
    \label{Eq:Geven}
    G_i^{\textrm{(even)}}\left(\vartheta,\chi\right) &:=& \cos^{2s} (i) = \left(\frac{\lambda_z}{\lambda}\right)^{2s} = (\sin\vartheta \sin\chi)^{2s}, \\
    \label{Eq:Grot}
    G_{i/2}^{\textrm{(rot)}}\left(\vartheta,\chi\right) &:=& \cos^{2s} (i/2) = \frac{1+s}{1+2s}\frac{1}{2^s} \left(1+\frac{\lambda_z}{\lambda}\right)^s = \frac{1+s}{1+2s} \frac{1}{2^s} \left(1+\sin\vartheta\sin\chi\right)^s,
\end{eqnarray}
where $s\geq 0$ is a constant. Here the parameter $k$ from the energy polytropic ansatz and the parameter $s$ from the inclination angle models are related each other by the condition $2k > s+ 7$ in order to guarantee a finite total number of particles, energy, and angular momentum~\cite{cGoS2023b}. We also have introduced the angle $\chi$ defined by $\displaystyle \lambda_z = \lambda \sin\vartheta \sin\chi$. For the bound orbits, the (dimensionless) parameter space $(\varepsilon, \lambda, \chi)$ has domain 
$\varepsilon_{\textrm{c}}(\xi) < \varepsilon \leq 1, \lambda_{\textrm{c}}(\varepsilon) \leq \lambda \leq \lambda_{\textrm{max}}(\varepsilon,\xi), 0 \leq \chi \leq 2\pi$, where $\varepsilon_{\textrm{c}}$ is the minimum energy at dimensionless radius $\xi$, see~\cite{cGoS2023b,cGoS2022,cGrR2025}. $\lambda_{\textrm{c}}$ is the critical value for the total angular momentum, and $\lambda_{\textrm{max}}$ is the maximum angular momentum permitted at dimensionless energy $\varepsilon$ and radius $\xi$ (see~\cite{pRoS17a,pRoS17b} and references therein). The explicit representation of these functions, are
\begin{equation}
    \varepsilon_{\textrm{c}}(r) = \left\{
    \begin{array}{lcl}
    \displaystyle \frac{\xi-2}{\sqrt{\xi\left(\xi-3\right)}}, & \hbox{for} & 4 \leq \xi \leq 6, \\ 
    & & \\
    \displaystyle \frac{\xi + 2}{\sqrt{\xi\left(\xi + 6\right)}}, & \hbox{for} & \xi \geq 6,
    \end{array}
    \right.
    \label{Eq:Ec}
\end{equation}
and
\begin{equation}
    \label{Eq:LcLmax}
    \lambda_{\textrm{c}}(\varepsilon) = \frac{4\sqrt{2}}{\sqrt{36 \varepsilon^2 - 8 - 27 \varepsilon^4 + \varepsilon\left( 9\varepsilon^2 - 8 \right)^{3/2}}}, \qquad 
    \lambda_{\textrm{max}}(\varepsilon,\xi) = \xi\sqrt{\frac{\varepsilon^2}{N(\xi)} - 1}.
\end{equation}

Using the one-particle DF as in~(\ref{Eq:OneParticleDistrFunction}) and rewriting the Lorentz-invariant volume form~(\ref{Eq:VolumeForm}) in terms of the dimensionless parameter space $(\varepsilon, \lambda, \chi)$, the entropy density covector field~(\ref{Eq:Smu}) yields the following:
\begin{eqnarray}
    \label{Eq:SmuExplicit}
    S_{\hat{\mu}}(x) &=& -\frac{k_B m^2}{\xi^2} \sum\limits_{\epsilon_r,\epsilon_\vartheta = \pm 1} \int\limits_{\varepsilon_{\textrm{c}}(\xi)}^1  \int\limits_{\lambda_{\textrm{c}}(\varepsilon)}^{\lambda_{\textrm{max}}(\varepsilon,\xi)} \int\limits_0^{2\pi} F_0(\varepsilon) G(i) \log\left[A F_0(\varepsilon) G(i) \right] p_{\hat{\mu}}(\epsilon_r,\epsilon_\vartheta)  \frac{d\varepsilon \: \lambda d\lambda \: d\chi}{\sqrt{\varepsilon^2 - U_{\lambda}(\xi)}}, \nonumber \\
    &=& -\frac{k_B m^2}{\xi^2} \sum\limits_{\epsilon_r,\epsilon_\vartheta = \pm 1} \left\{ \int\limits_{\varepsilon_{\textrm{c}}(\xi)}^1 d\varepsilon F_0(\varepsilon) \log\left[ A F_0(\varepsilon) \right] \int\limits_{\lambda_{\textrm{c}}(\varepsilon)}^{\lambda_{\textrm{max}}(\varepsilon,\xi)}  \frac{\lambda \: d\lambda}{\sqrt{\varepsilon^2 - U_{\lambda}(\xi)}} \int\limits_0^{2\pi} d\chi G(i) p_{\hat{\mu}}(\epsilon_r,\epsilon_\vartheta) \right. \nonumber\\
    & & \qquad\qquad\qquad\qquad \left. + \int\limits_{\varepsilon_{\textrm{c}}(\xi)}^1 d\varepsilon F_0(\varepsilon) \int\limits_{\lambda_{\textrm{c}}(\varepsilon)}^{\lambda_{\textrm{max}}(\varepsilon,\xi)}  \frac{\lambda \: d\lambda}{\sqrt{\varepsilon^2 - U_{\lambda}(\xi)}} \int\limits_0^{2\pi} d\chi G(i) \log G(i) p_{\hat{\mu}}(\epsilon_r,\epsilon_\vartheta) \right\}.
\end{eqnarray}
Here, we have used the expansion of the covector field $p$ in terms of an orthonormal basis $p = p_{\hat{\mu}} e^{\hat{\mu}}$ with components,
\begin{equation}
    \label{Eq:Orthonormalbasis}
    p_{\hat{\mu}} = m\left(-\frac{\varepsilon}{\sqrt{N(\xi)}}, \frac{\epsilon_r}{\sqrt{N(\xi)}} \sqrt{\varepsilon^2 - U_{\lambda}(\xi)}, \frac{\epsilon_\vartheta}{\xi}\sqrt{\lambda^2 - \frac{\lambda_z^2}{\sin^2\vartheta}}, \frac{\lambda_z}{\xi\sin\vartheta} \right).
\end{equation}
and the signs $\epsilon_r = \epsilon_\vartheta =\pm 1$ determine the corresponding signs of $p_{\hat{1}}$ and $p_{\hat{2}}$, and the orthonormal basis of vector fields are given by $\displaystyle e^{\hat{\mu}} = \left(-N(\xi)^{-1/2}, N(\xi)^{1/2}, \xi^{-1}, (\xi\sin\vartheta)^{-1} \right)$. 

The resulting nonvanishing components of the entropy flux covector field~(\ref{Eq:SmuExplicit}) for the non-rotating~(\ref{Eq:Geven}) and rotating~\ref{Eq:Grot}) models are given by
\begin{eqnarray}
    \label{Eq:S0}
    S^{(\textrm{even},\textrm{rot})}_{\hat{0}}(x) &=& \frac{2 k_B m^3}{\xi N(\xi)} \left\{ \mathcal{I}^{(\textrm{even},\textrm{rot})}_{i,i/2}(\vartheta) \int\limits_{\varepsilon_{\textrm{c}}(\xi)}^1 d\varepsilon \varepsilon F_0(\varepsilon) \log\left[A F_0(\varepsilon)\right] \lambda_{\textrm{max}}(\varepsilon, \xi) \sqrt{1 - b(\varepsilon, \xi)^2}, \right. \nonumber\\
    & & \left.\quad \qquad \qquad + \widetilde{\mathcal{I}}^{(\textrm{even},\textrm{rot})}_{i,i/2}(\vartheta) \int\limits_{\varepsilon_{\textrm{c}}(\xi)}^1 d\varepsilon \varepsilon F_0(\varepsilon) \lambda_{\textrm{max}}(\varepsilon, \xi) \sqrt{1 - b(\varepsilon, \xi)^2} \right\}, \\
    \label{Eq:S3}
    S^{(\textrm{rot})}_{\hat{3}}(x) &=& \frac{k_B m^3}{\xi^2 \sqrt{N(\xi)}} \left\{ \widehat{\mathcal{I}}^{(\textrm{rot})}_{i/2}(\vartheta) \int\limits_{\varepsilon_{\textrm{c}}(\xi)}^1 d\varepsilon F_0(\varepsilon) \log\left[A F_0(\varepsilon)\right] \lambda^2_{\textrm{max}}(\varepsilon, \xi) B(\varepsilon, \xi) \right. \nonumber\\
    & & \left.\quad \qquad \qquad + \overline{\mathcal{I}}^{(\textrm{rot})}_{i/2}(\vartheta) \int\limits_{\varepsilon_{\textrm{c}}(\xi)}^1 d\varepsilon F_0(\varepsilon) \lambda^2_{\textrm{max}}(\varepsilon, \xi) B(\varepsilon, \xi) \right\},     
\end{eqnarray}
where we have introduced the shorthand notation $b$ and $B$, which are given by
\begin{equation}
    \label{Eq:bB}
    b(\varepsilon, \xi) := \frac{\lambda_{\textrm{c}}(\varepsilon)}{\lambda_{\textrm{max}}(\varepsilon, \xi)}, \quad \hbox{and} \quad 
    B(\varepsilon, \xi) := \frac{\pi}{2} + b(\varepsilon, \xi)\sqrt{1-b(\varepsilon, \xi)^2} - \arctan\frac{b(\varepsilon, \xi)}{\sqrt{1-b(\varepsilon, \xi)^2}},
\end{equation}
and the explicit expressions for the angular functions $\displaystyle \mathcal{I}^{(\textrm{even},\textrm{rot})}_{i,i/2}(\vartheta)$, $\displaystyle \widetilde{\mathcal{I}}^{(\textrm{even},\textrm{rot})}_{i,i/2}(\vartheta)$, $\displaystyle \widehat{\mathcal{I}}^{(\textrm{rot})}_{i/2}(\vartheta)$, and $\displaystyle \overline{\mathcal{I}}^{(\textrm{rot})}_{i/2}(\vartheta)$ are provided in the Appendix~\ref{Appx:A}.

The invariant entropy density~(\ref{Eq:EntropyDensity}) for the even model~(\ref{Eq:Geven}) is given by 
\begin{equation}
    \label{Eq:Seven}
    S^{(\textrm{even})}(x) = S_{\hat{0}}^{(\textrm{even})}(x),
\end{equation}
while for the rotating model~(\ref{Eq:Grot}), it yields
\begin{equation}
    \label{Eq:Srot}
    S^{(\textrm{rot})}(x) = \sqrt{ \left[ S^{(\textrm{rot})}_{\hat{0}}(x) \right]^2 - \left[ S^{(\textrm{rot})}_{\hat{3}}(x) \right]^2}. 
\end{equation}

\begin{figure}[h!]
\centerline{
\includegraphics[scale=0.215]{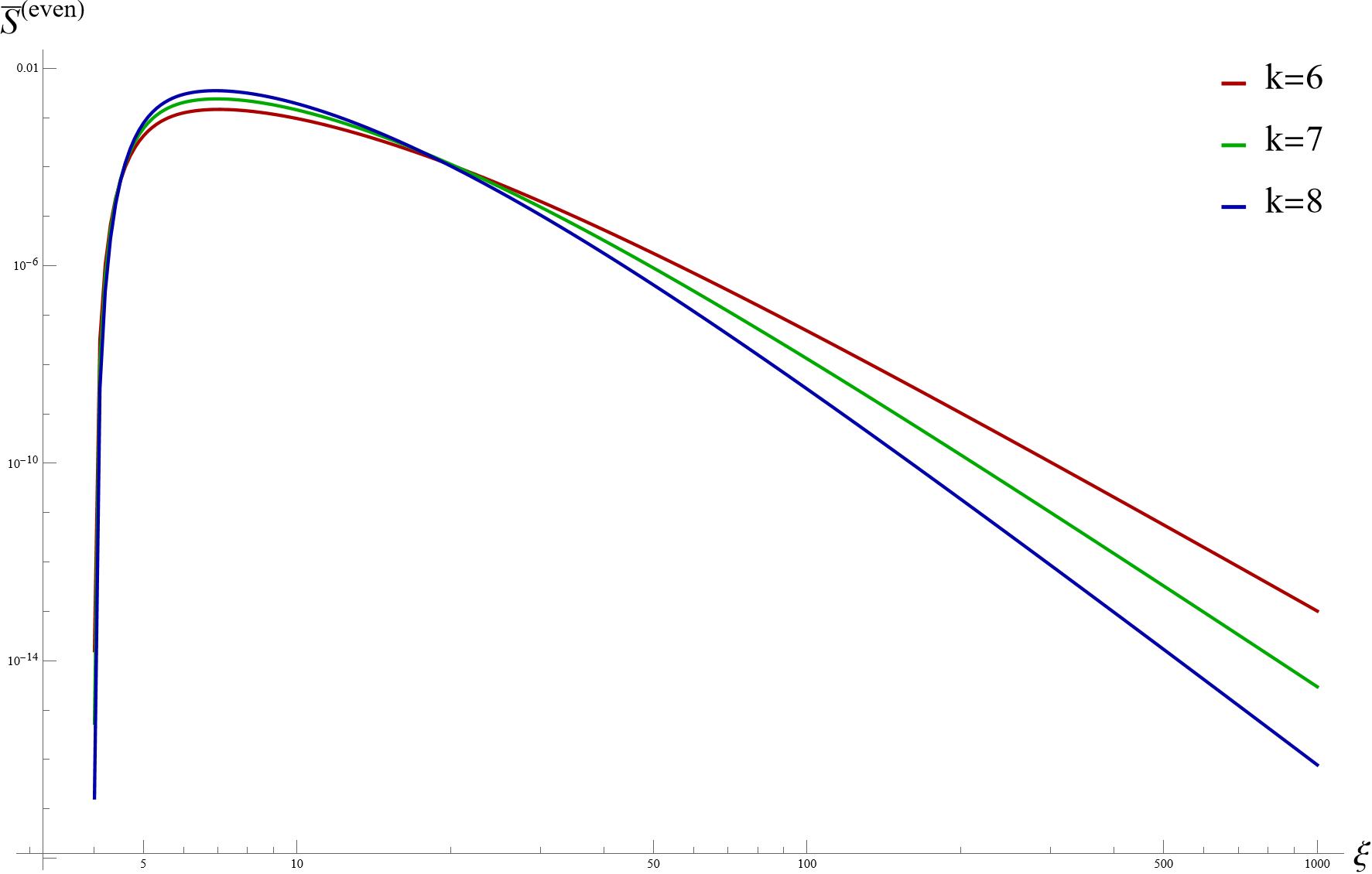}
\includegraphics[scale=0.215]{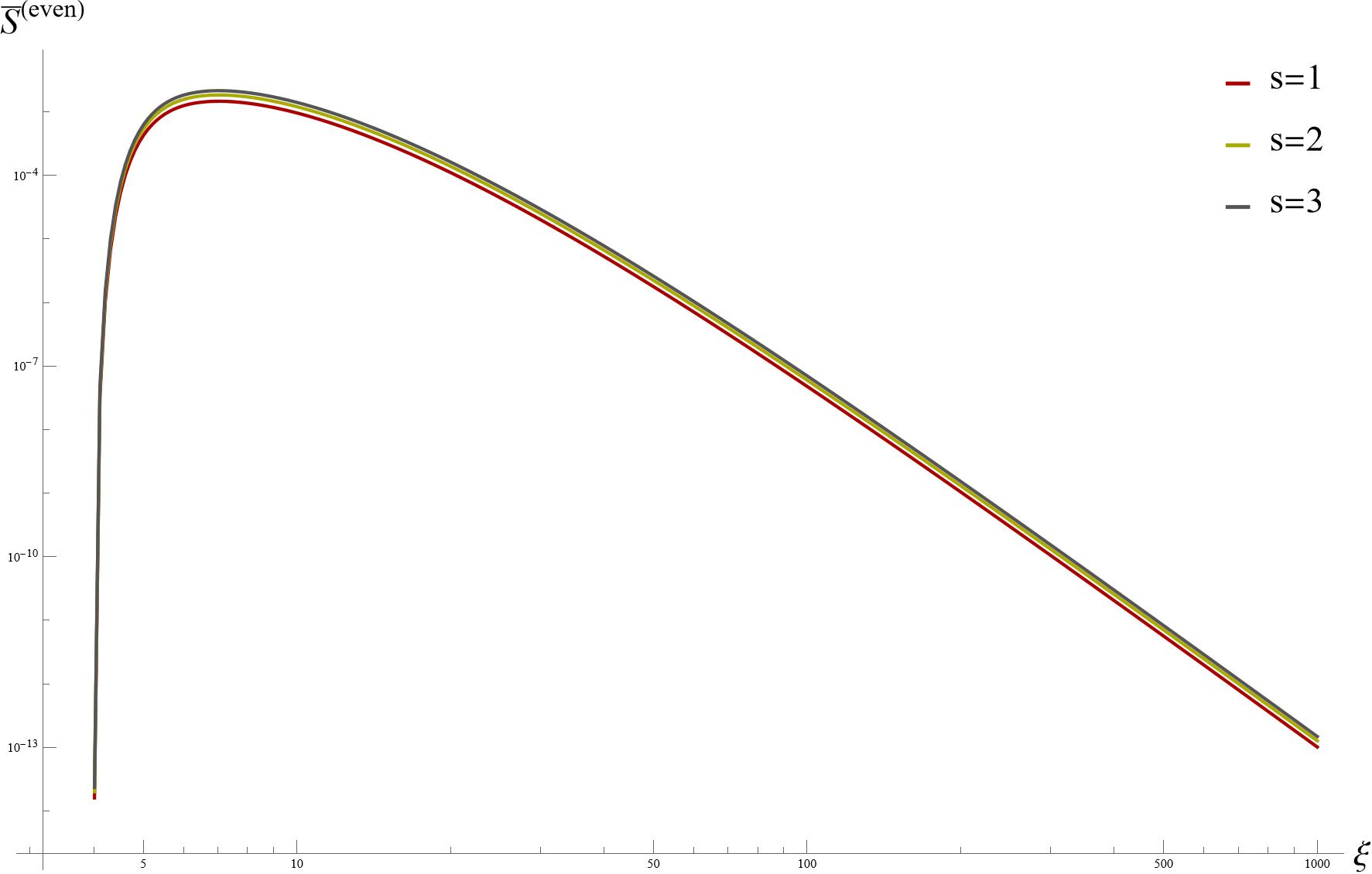}}
\caption{Log-log plot showing the behavior of the normalized entropy density as a function of the dimensionless areal radius $\xi$ in the equatorial plane in the non-rotating model. Left panel: plot for different parameter values of $k=6,7,8$ and $(s,\varepsilon_0)=(1,1)$. Right panel: plot for different parameter values of $s=1,2,3$ and $(k,\varepsilon_0)=(6,1)$.}
\label{Fig:Seven}
\end{figure}

These macroscopic observables depend on the amplitude $\alpha$ parameter through the function $F_0$~(\ref{Eq:F0}); consequently, it is necessary to eliminate this dependence. While previous works~\cite{pRoS17a,pRoS17b} fixed this parameter at the infinity of the configuration, in the present study, we absorb this dependency into the total particle number of the gas configurations, following the approach in~\cite{cGoS2023b,cGrR2025}, which is a conserved quantity. This total particle number can be expressed in the action-angle variables, yielding a more streamlined expression since it $\mathcal{F}$ depends only on conserved quantities.  Switching to variables analogous to the semi-latus rectum and eccentricity simplifies the numerical integration by mapping the system onto a simpler domain~\cite{pRoS18,pRoS22}. The final expression 
\begin{equation}
    \label{Eq:TotalNumberParticlesep}
    \frac{\mathcal{N}_{\textrm{gas}}}{M^3 m^3 \alpha} = \frac{8\pi^2}{2s+1} \int\limits_{0}^1 de \int\limits_{6+2e}^{\infty} \left(1-\frac{\varepsilon(e,p)}{\varepsilon_0} \right)^{k-\frac{3}{2}}_+ \left(\mathbb{H}_2(e,p)-\mathbb{H}_0(e,p)\right) \dfrac{e\left((p-6)^2-4e^2\right)}{(p-e^2-3)^3}dp,
\end{equation}
is computed numerically using~\cite{Mathematica}. For more details of this result, we refer to~\cite{cGoS2022,cGoS2023b,cGrR2025}. However, this is not the only dependency involved; the constants $A$ and $\alpha$ also appear within the logarithmic functions accompanying $F_0$. To decouple these constants, we introduce a rescaling such that $A\alpha=1$ in all subsequent results. This simplification is justified as the constants within the logarithmic argument do not alter the functional behavior or shape of the observables; they merely introduce a vertical shift of $\log A\alpha$ to the resulting plots.
We start the analysis for the normalized profile of invariant entropy density for the models (even-rot) which is given by
\begin{equation}
    \label{Eq:BarS}
    \bar{S}(\xi,\vartheta) = \frac{M^3}{k_B \mathcal{N}_{\text{gas}}}S(\xi,\vartheta).
\end{equation}
for different choices of the parameter space. The figure~\ref{Fig:Seven} (left and right) shows the behavior of the normalized entropy density as a function of the dimensionless areal radius for a non-rotating kinetic gas. One can notice from this plot that for larger values of the parameter $k$ while keeping constant the parameter $s$, the configuration reaches maximum levels of entropy density at radii near the BH due to the polytropic ansatz contribution to the DF, while the configuration decays at long radii, as expected. This behavior is observed in other related works~\cite{cGrR2025}. In contrast, by varying $s$ and keeping $k$ constant, one can notice that for greater values of $s$, the entropy density is more concentrated at radii nearer to the black hole.

\begin{figure}[h!]
\centerline{
\includegraphics[scale=0.225]{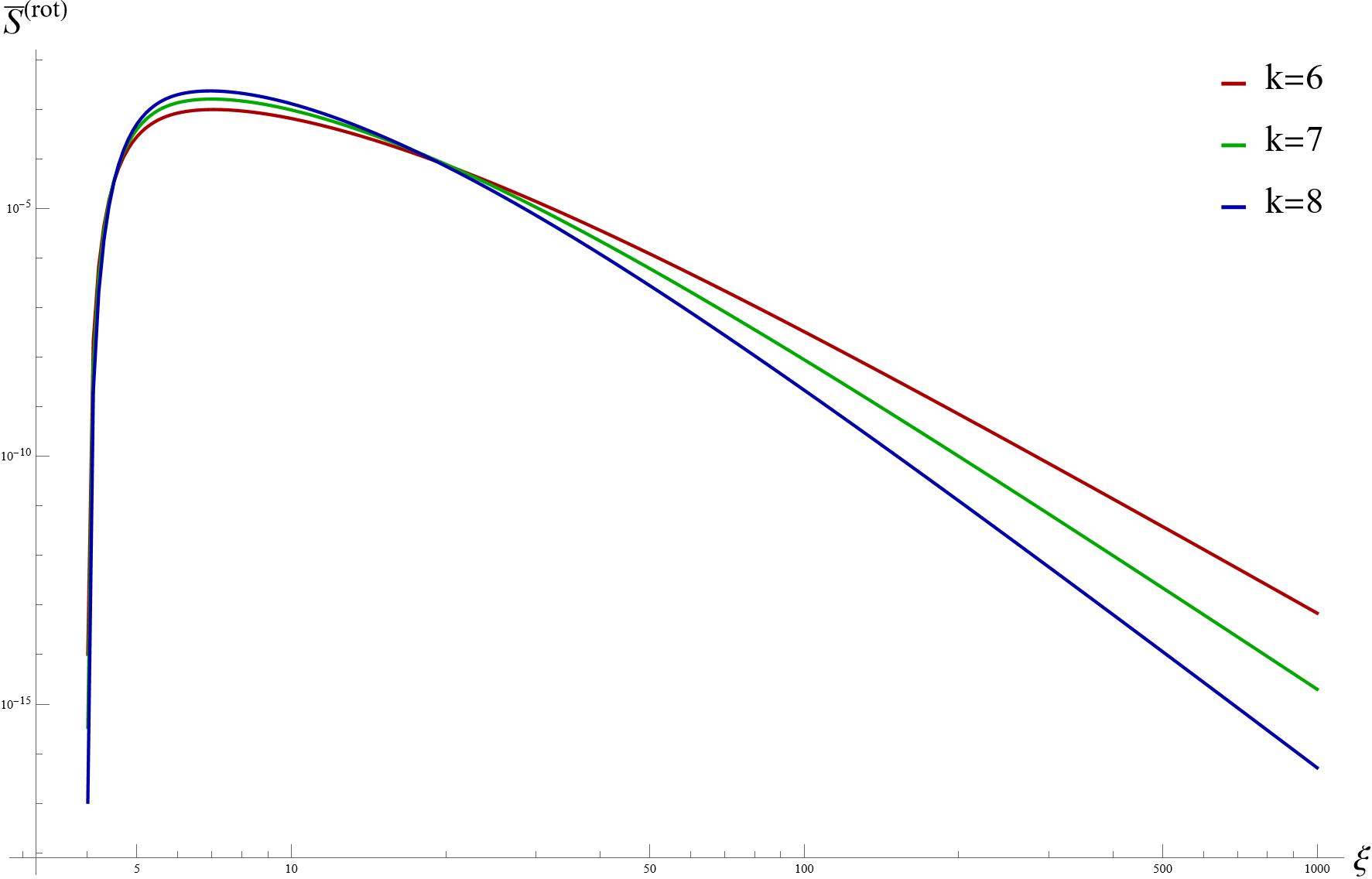}
\includegraphics[scale=0.225]{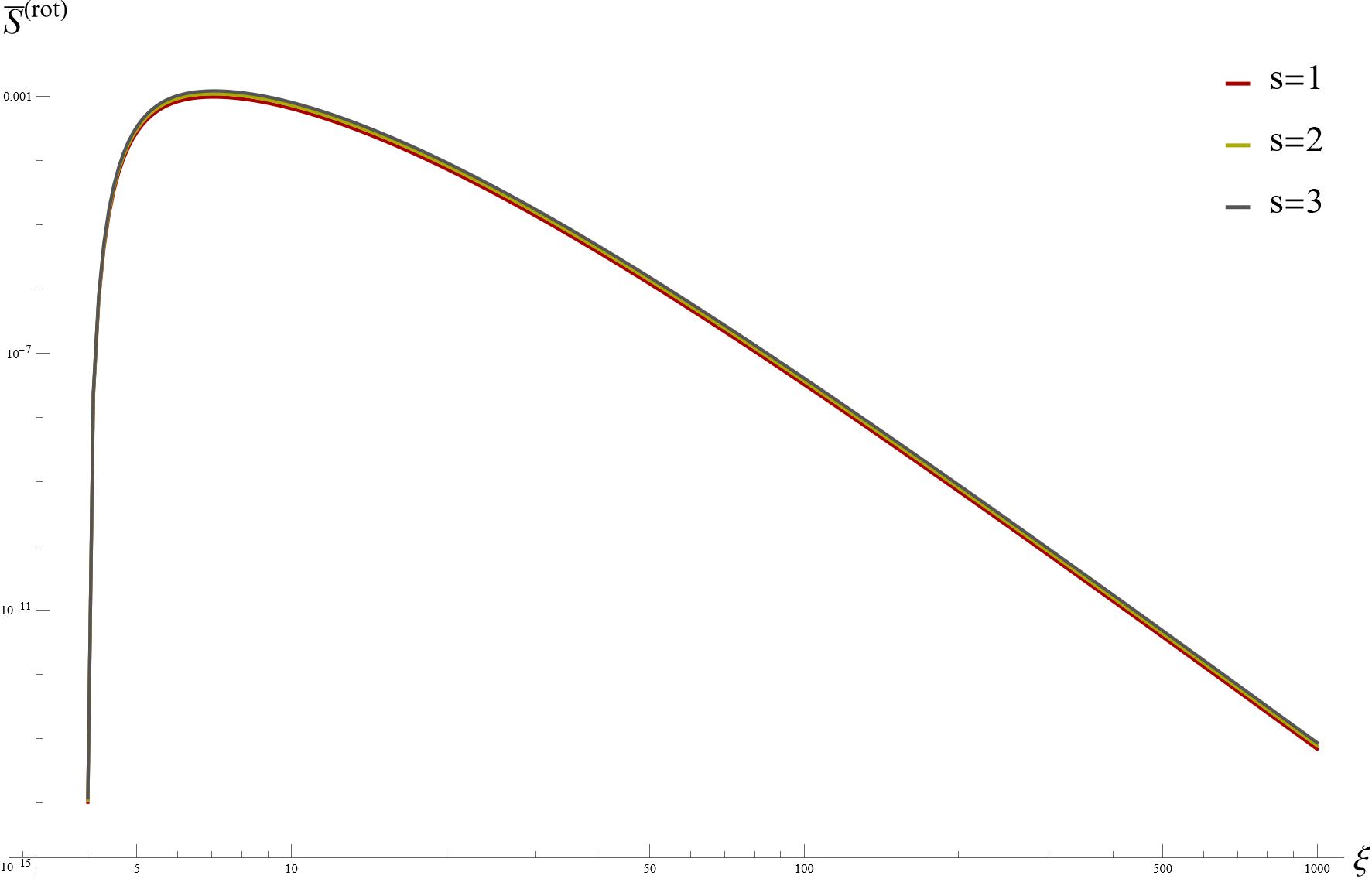}}
\caption{Log-log plot showing the behavior of the normalized entropy density as a function of the dimensionless areal radius $\xi$ in the equatorial plane in the rotating model. Left panel: plot for different parameter values of $k=6,7,8$ and $(s,\varepsilon_0)=(1,1)$. Right panel: plot for different parameter values of $s=1,2,3$ and $(k,\varepsilon_0)=(6,1)$.}
\label{Fig:Srot}
\end{figure}

The behavior of $S$ when varying the parameters $k$ and $s$ for a rotating gas is the same as for the non-rotating gas, as shown in figure~\ref{Fig:Srot} (left and right).

Comparing the entropy between the two models through the ratio $\displaystyle \bar{S}^{(\textrm{even})}/\bar{S}^{(\textrm{rot})}$, we can make significant distinctions between the models, given different values of $k$ and for different regions around the BH. For all explored values of the polytropic index parameter $k$, $\displaystyle \bar{S}^{(\textrm{even})}/\bar{S}^{(\textrm{rot})}>1$. This behavior is expected, as (\ref{Eq:Srot}) yields a result from the difference between the two non-vanishing components of the entropy flux covector, while $S^{(\textrm{even})}$ is directly proportional two the only non-vanishing term of the entropy flux for the non-rotating gas. Though the ratio of the entropy between the two models of gas always remains above the unity, the effect of the non-zero total angular momentum induced by the inclination angle model (\ref{Eq:Grot}) implies a notorious reduction in entropy density for said model. Also, one can notice that, for any value of $k$, the ratio presents a minimum for radii nearer to the BH. At this region, we expect that the entropy flux distributes in a less efficient way for the rotating gas than for its non-rotating counterpart. This suggests the system presents a greater sensitivity to rotational anisotropy due to the asymmetry in the total angular momentum induced by the \textit{rot} inclination angle model. Beyond this region, the ratio increases toward larger radii. This may indicate that the entropy reduction induced by the rotation of the configuration of gas is not exclusively an effect of the strong-gravity of the background, but rather a consequence of the behavior of the system due to space-phase mixing.

\begin{figure}[h!]
\centerline{
\includegraphics[scale=0.25]{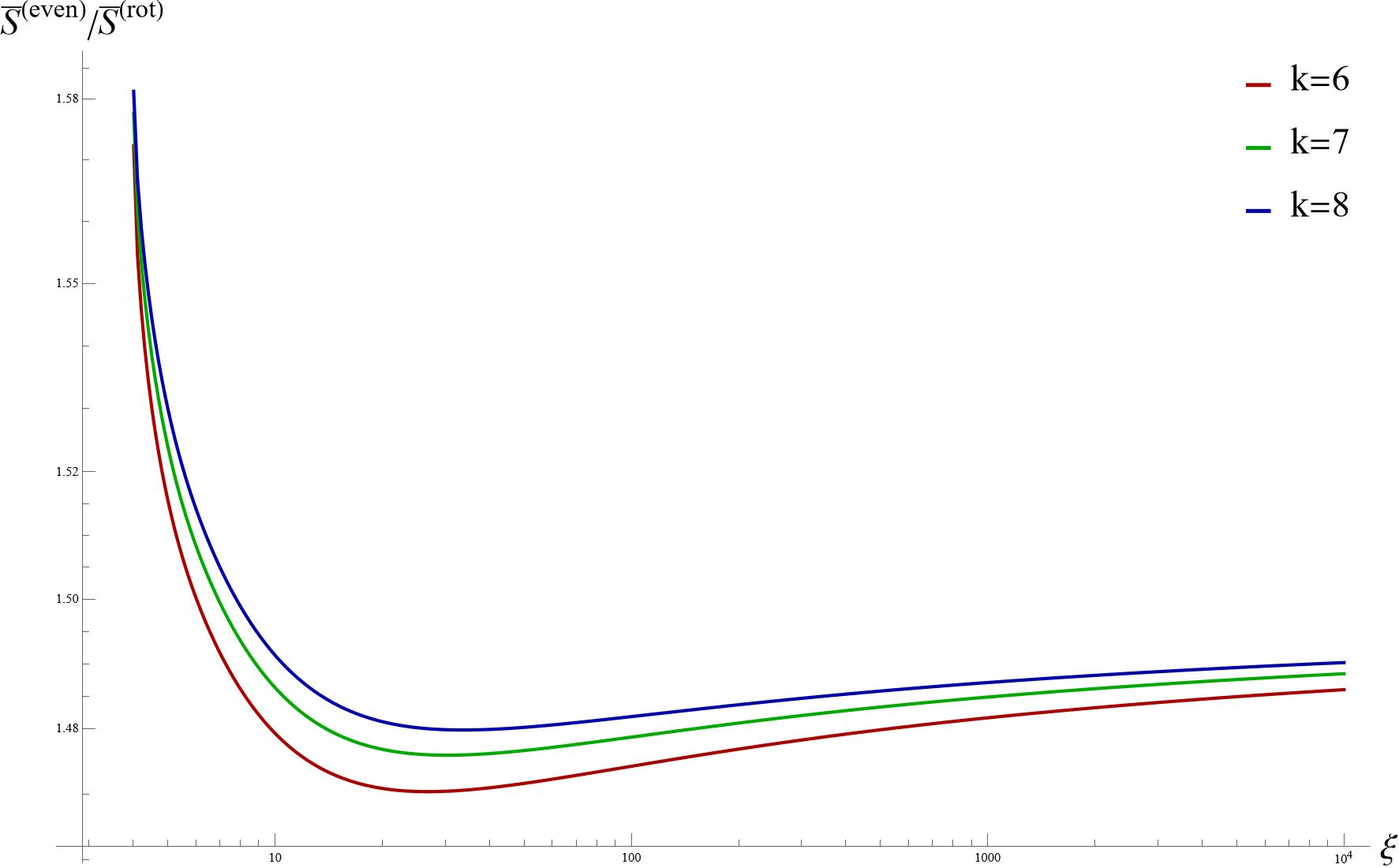}}
\caption{Logarithmic plot of the relative difference with the normalized entropy density as a function of the dimensionless area radius $\xi$ in the equatorial plane for $k=6,7,8$ and $(s,\varepsilon_0)=(1,1)$. The relative difference between the even and rotating models for different parameter space choices reveals a relativistic effect on the entropy density profiles invariant between the non-rotating and rotating models. The maximum relative difference occurs at the inner radius of the disk, while for intermediate radii it decreases to a minimum and then increases monotonically for large radii.}
\label{Fig:SevenSrot}
\end{figure}

\section{Principal pressures and anisotropy parameter behavior}
\label{Sec:Anisotropy}

As mentioned previously in Section~\ref{Sec:Preliminaries}, from the energy-momentum-stress tensor~(\ref{Eq:Tmunu}) we can obtain the principal pressures of the system, which are macroscopic observables of interest since they provide information about the gas' behavior in terms of the velocity dispersion. Furthermore, the relationship between these principal pressures allows us to define an anisotropy parameter that quantifies how the random motions differ from the radial to the tangential direction, thus allowing us to infer a bias in the radial or tangential contributions over the kinetic gas.

\begin{figure}[h!]
\centerline{
\includegraphics[scale=0.3]{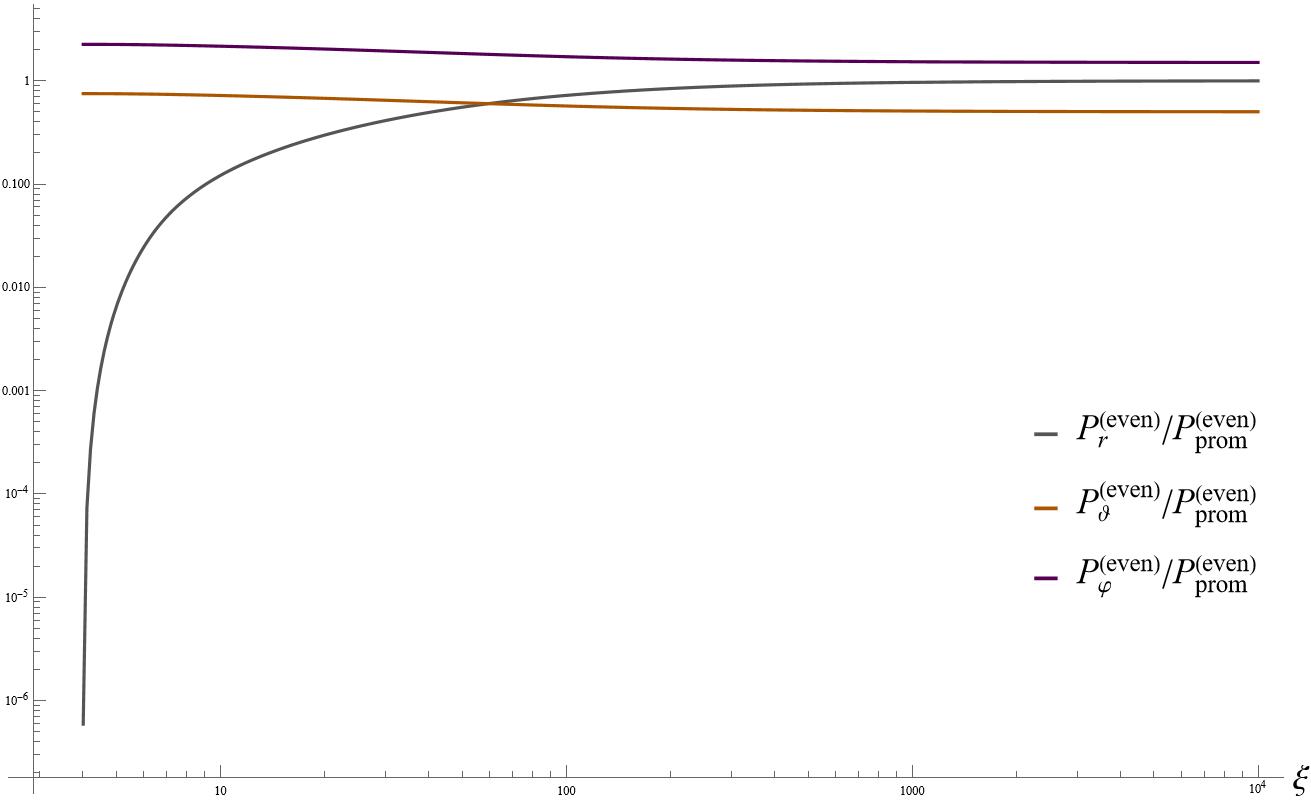}
\includegraphics[scale=0.3]{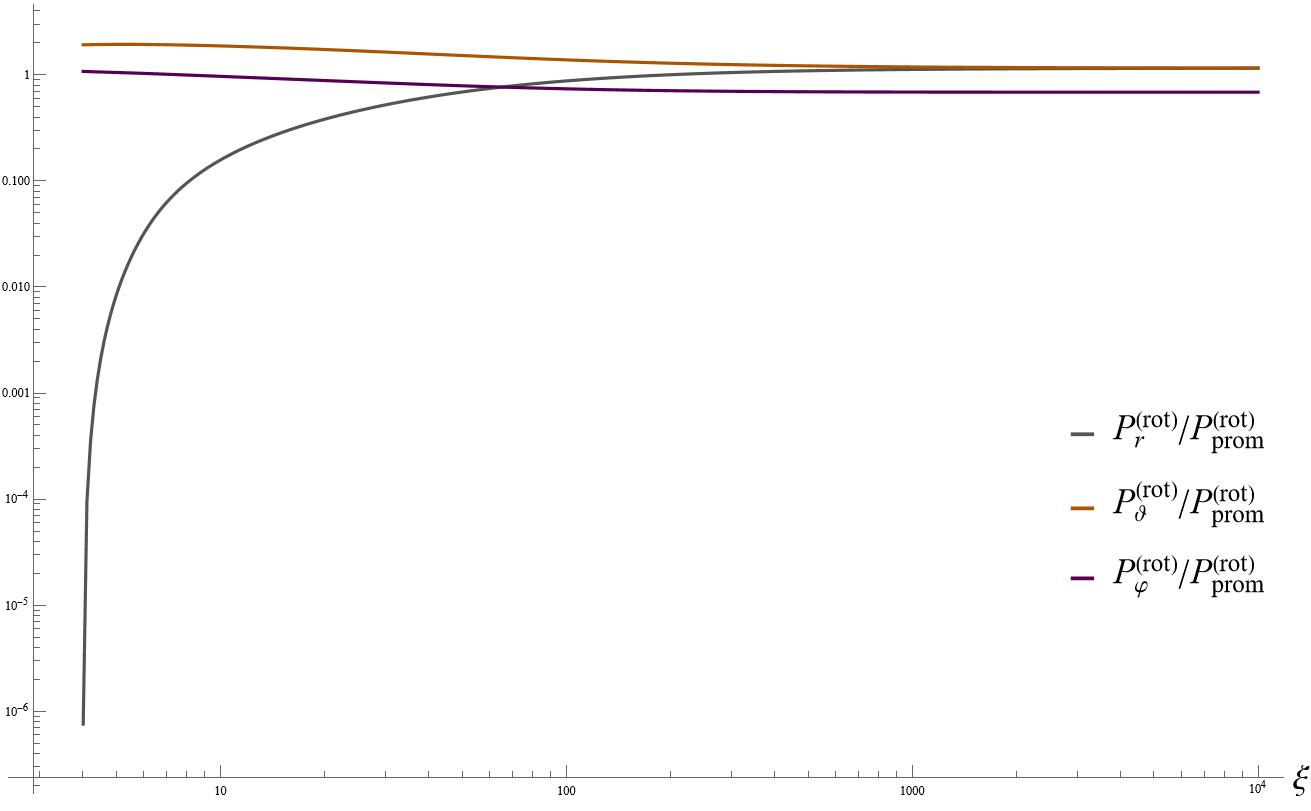}}
\caption{Log–log plot showing the behavior of the principal pressures (normalized by the average pressure $P_{\textrm{prom}}$) as a function of the dimensionless radius $\xi$ in the equatorial plane. For both panels (left and right), the parameter values are $(k,s,\varepsilon_0)=(6,1,1)$. Left panel: plot for the non-rotating gas model. Right panel: plot for the rotating gas model. Note that in these cases the three pressures are different from each other, with the radial pressure converging to polar pressure for large $\xi$ in both cases.}
\label{Fig:Pevenrot}
\end{figure}

The components of the pressure tensor for the non-rotating model are summarized as follows:
\begin{equation}
    P^{(\textrm{even})}_{\hat{r}}(x) = T^{\hat{1}}{}_{\hat{1}}^{(\textrm{even})}(x), \qquad 
    P^{(\textrm{even})}_{\hat{\vartheta}}(x) = T^{\hat{2}}{}_{\hat{2}}^{(\textrm{even})}(x), \qquad \hbox{and} \qquad
    P^{(\textrm{even})}_{\hat{\varphi}}(x) = T^{\hat{3}}{}_{\hat{3}}^{(\textrm{even})}(x),
    \label{Eq:Peven}
\end{equation}
and the associated components for the rotating model are:
\begin{eqnarray}
    P^{(\textrm{rot})}_{\hat{r}}(x) &=& T^{(\textrm{rot})}_{\hat{1}\hat{1}}(x), \qquad 
    P^{(\textrm{rot})}_{\hat{\vartheta}}(x) = T^{(\textrm{rot})}_{\hat{2}\hat{2}}(x), \qquad \hbox{and} \qquad \nonumber\\
    P^{(\textrm{rot})}_{\hat{\varphi}}(x) &=& \frac{1}{2}\left[ -T^{(\textrm{rot})}_{\hat{0}\hat{0}}(x) + T^{(\textrm{rot})}_{\hat{3}\hat{3}}(x) + \sqrt{ \left(T^{(\textrm{rot})}_{\hat{3}\hat{3}}(x) + T^{(\textrm{rot})}_{\hat{0}\hat{0}}(x)\right)^2 - 4 \left(T^{(\textrm{rot})}_{\hat{0}\hat{3}}(x)\right)^2 }\right].
    \label{Eq:Prot}
\end{eqnarray}
First, we will focus on the behavior of principal pressures, in which one can introduce the average pressure as follows
\begin{equation}
    P^{(\textrm{even,rot})}_{\textrm{prom}} := \frac{1}{3}\left(P^{(\textrm{even,rot})}_{\hat{r}} + P^{(\textrm{even,rot})}_{\hat{\vartheta}} + P^{(\textrm{even,rot})}_{\hat{\varphi}}\right)
    \label{Eq:Pprom}
\end{equation}
for both models (even) or (rot) correspondingly. Figure~\ref{Fig:Pevenrot} shows the principal pressures~(\ref{Eq:Peven}-\ref{Eq:Prot}) as functions of the dimensionless coordinate $\xi$ on the equatorial plane for the non-rotating and rotating models (left and right). It can be noticed that in both models these pressures are different from each other in the region with radii near the BH, while for larger radii, the principal pressures on the radial and polar directions are always equal to each other. This can be understood by realizing that the explicit expressions of~(\ref{Eq:Peven}-\ref{Eq:Prot}) are different from each other; however, for large values of $\xi$, the results agree with the non-relativistic limit discussed for similar models in~\cite{cGoS2023a}. On the other hand, the polar and azimuthal pressures are related by a factor depending on the $s$-parameter in the non-rotating case, while for the rotating case these principal pressures are related by an expression with a complex dependence on the polar angle; see equations (36-39) and (42-46) in~\cite{cGrR2025}. The behavior on the equatorial plane is shown in figures for both models for different values of the parameters $(k,s,\varepsilon_0)$.

Second, following the steps of the classic stellar-dynamical framework provided by Binney \& Tremaine~\cite{BinneyTremaine-Book,lCh2018}, anisotropy quantifies how a system’s random stellar motions differ between the radial direction (toward or away from the center) and the tangential directions (around the center). One commonly used definition is
\begin{equation}
    \beta(r) = 1 - \frac{\sigma_t^2(r)}{2 \sigma_r^2(r)},
\end{equation}
where $\sigma_r$ and $\sigma_t$ are the one‐dimensional velocity dispersions in the radial and combined tangential directions, respectively. When $\beta=0$ the stars’ orbits have no preferred direction, they are perfectly isotropic; positive $\beta$ indicates radially biased orbits (more radial excursions), while negative $\beta$ signals tangentially biased orbits (more circular motion).

Intending to further explore the structure of these models with total (or zero) angular momentum, we introduce the anisotropy parameter as defined in~\cite{cGoS2023a} and in analogy to~\cite{BinneyTremaine-Book}, as follows:
\begin{equation}
    \label{Eq:Beta}
    \beta := 1 - \frac{P_{\perp}}{P_{\parallel}} \equiv 1 - \frac{P_{\hat{\vartheta}} + P_{\hat{\varphi}}}{2 P_{\hat{r}}},
\end{equation}
where the symbol $\perp$ denotes the total contribution in the perpendicular motion associated with angular and azimuthal components and $\parallel$ refers to the contribution in the radial motion. These are associated with the radial or tangential components of the pressure tensor. 
\begin{figure}[htbp]
    \centering
    \begin{subfigure}[b]{0.47\textwidth}
        \includegraphics[width=8.5cm]{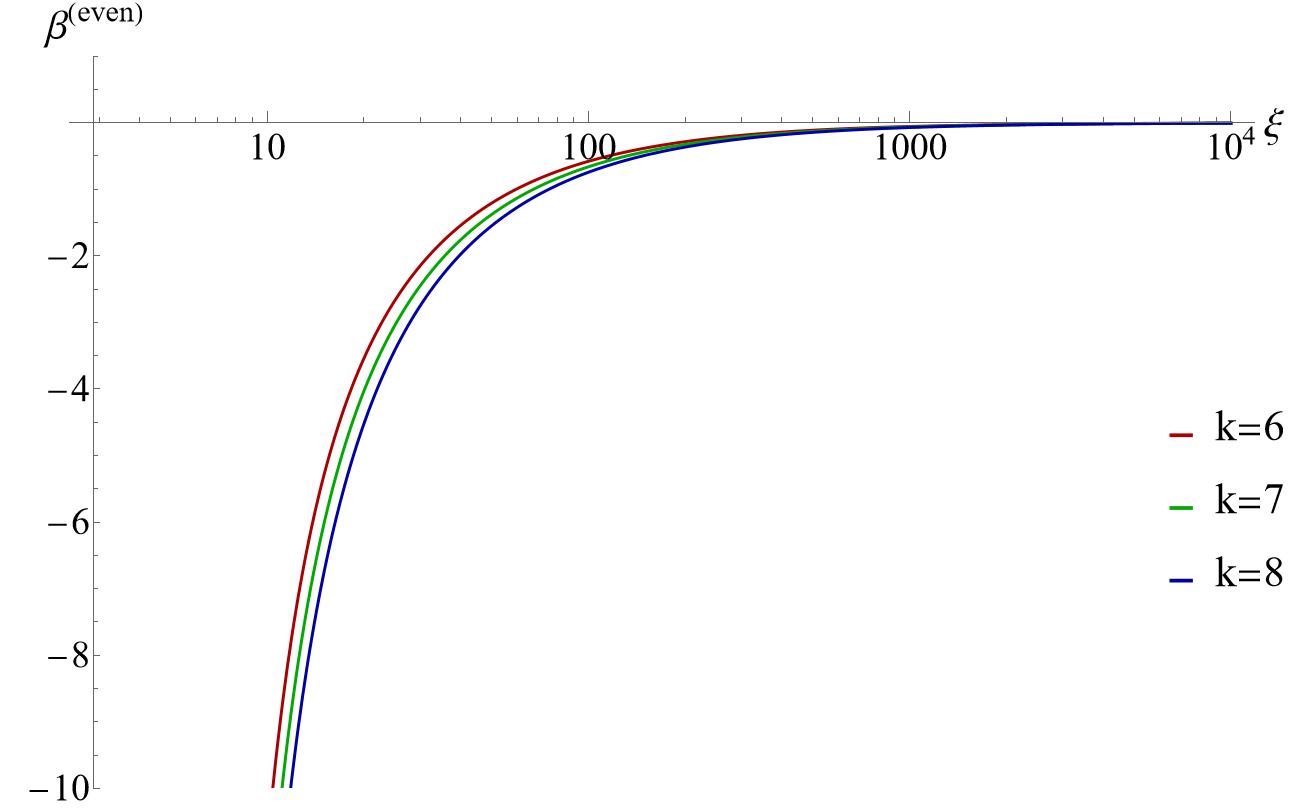}
    \end{subfigure}
    \hfill 
    \begin{subfigure}[b]{0.47\textwidth}
        \includegraphics[width=8.5cm]{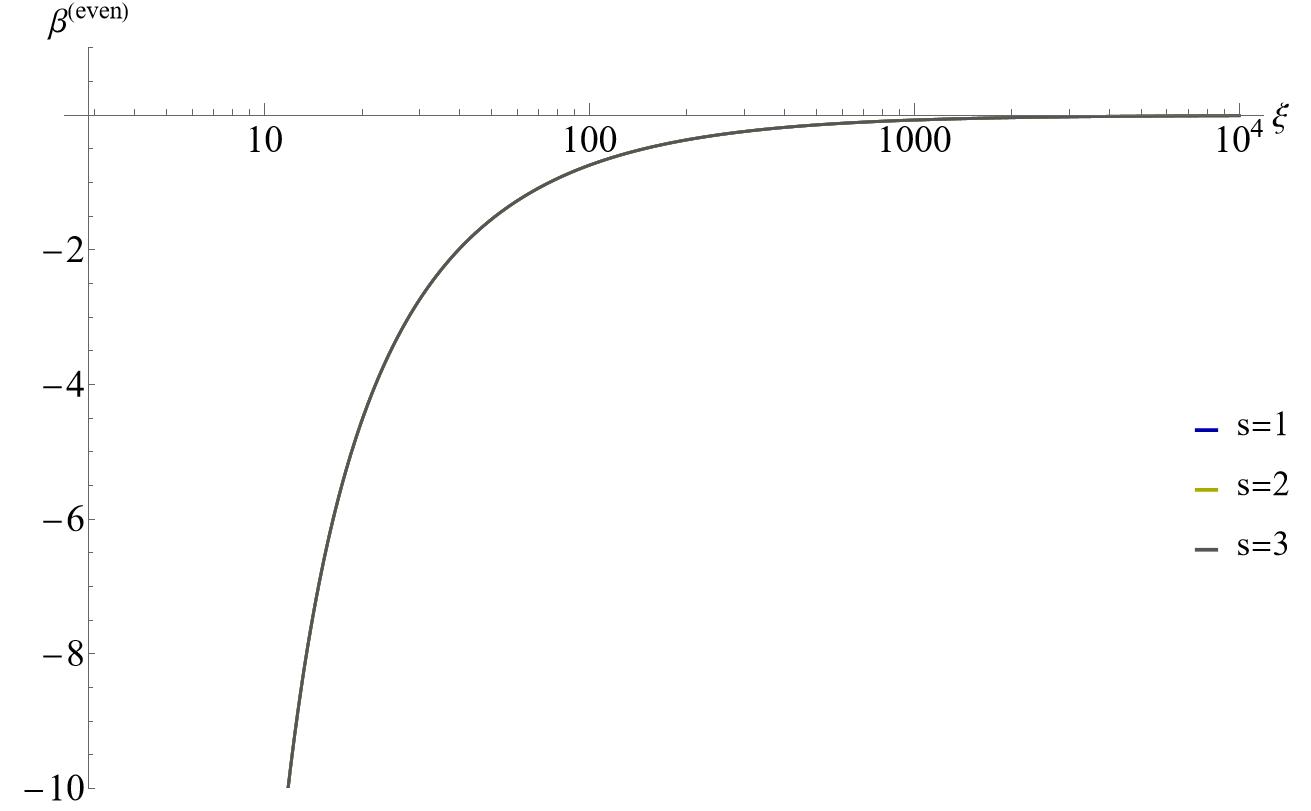}
    \end{subfigure}
    \caption{Log plot showing the behavior of the anisotropy parameter for the even model as a function of the dimensionless radius $\xi$ in the equatorial plane. Left panel: plot for the non-rotating gas model with parameter values $(s,\varepsilon_0)=(1,1)$ and different $k$ values. Right panel: plot for the non-rotating gas model with parameter values $(k,\varepsilon_0)=(6,1)$ and different $s$ values.}
    \label{Fig:Beven}
\end{figure}

Figure~\ref{Fig:Beven} shows the behavior of the anisotropy parameter as a function of the dimensionless radius $\xi$ in the equatorial plane for the non-rotating gas model (zero angular momentum). In the left panel, it can be noticed that the anisotropy parameter exhibits a significant dependence on the value of $k$, especially in the intermediate $\xi$ region. For a fixed $\xi$, the anisotropy decreases as $k$ increases, indicating that, even in the absence of rotation, the choice of the parameter $k$ modifies the structure of macroscopic observables in the non-collisional regime. In the right panel, unlike the case discussed before, the anisotropy parameter is practically independent of the value of $s$, exhibiting universal behavior for every $\xi$. This suggests that, within the same non-rotating model, the distribution function is insensitive to the internal parameterization of $s$, while it does respond to variations in $k$.

One can note that, in both configurations of non-rotating gas, the anisotropy parameter exhibits negative values across the entire range of $\xi$ considered, indicating a preferential direction in the gas velocity distribution. This behavior is consistent with the non-collisional nature of the system, where the distribution function, depending solely on the motion constants, can generate systematic deviations from an isotropic distribution. Furthermore, it is observed that for large radii, the anisotropy parameter tends asymptotically to zero in all cases. This limit indicates that, at sufficient distances from the center, the gas recovers isotropic behavior regardless of the values of the parameters $k$ and $s$. This trend is expected, since in external regions the gravitational field weakens and the particle trajectories become increasingly rectilinear and less influenced by the central geometry, leading to a more isotropic velocity distribution.
\begin{figure}[htbp]
    \centering
    \begin{subfigure}[b]{0.47\textwidth}
        \includegraphics[width=8.5cm]{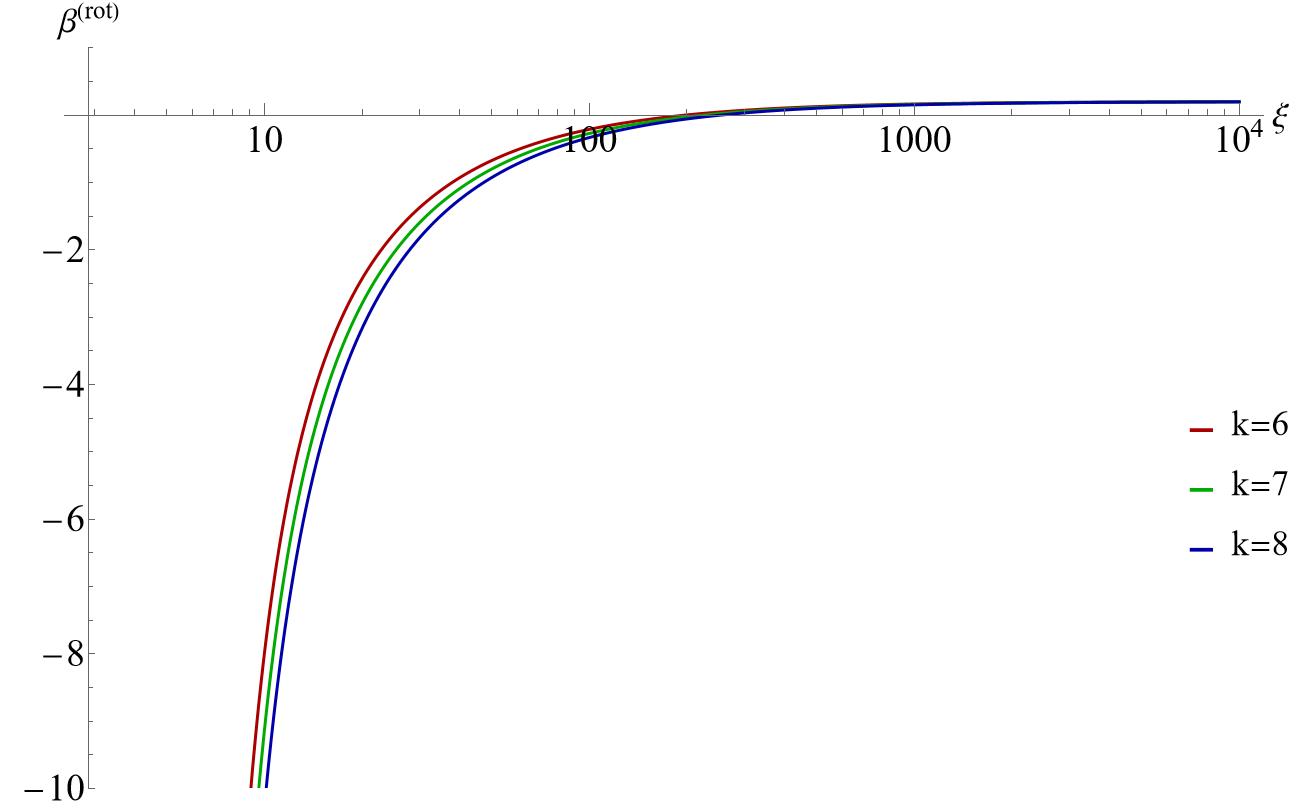}
    \end{subfigure}
    \hfill 
    \begin{subfigure}[b]{0.47\textwidth}
        \includegraphics[width=8.5cm]{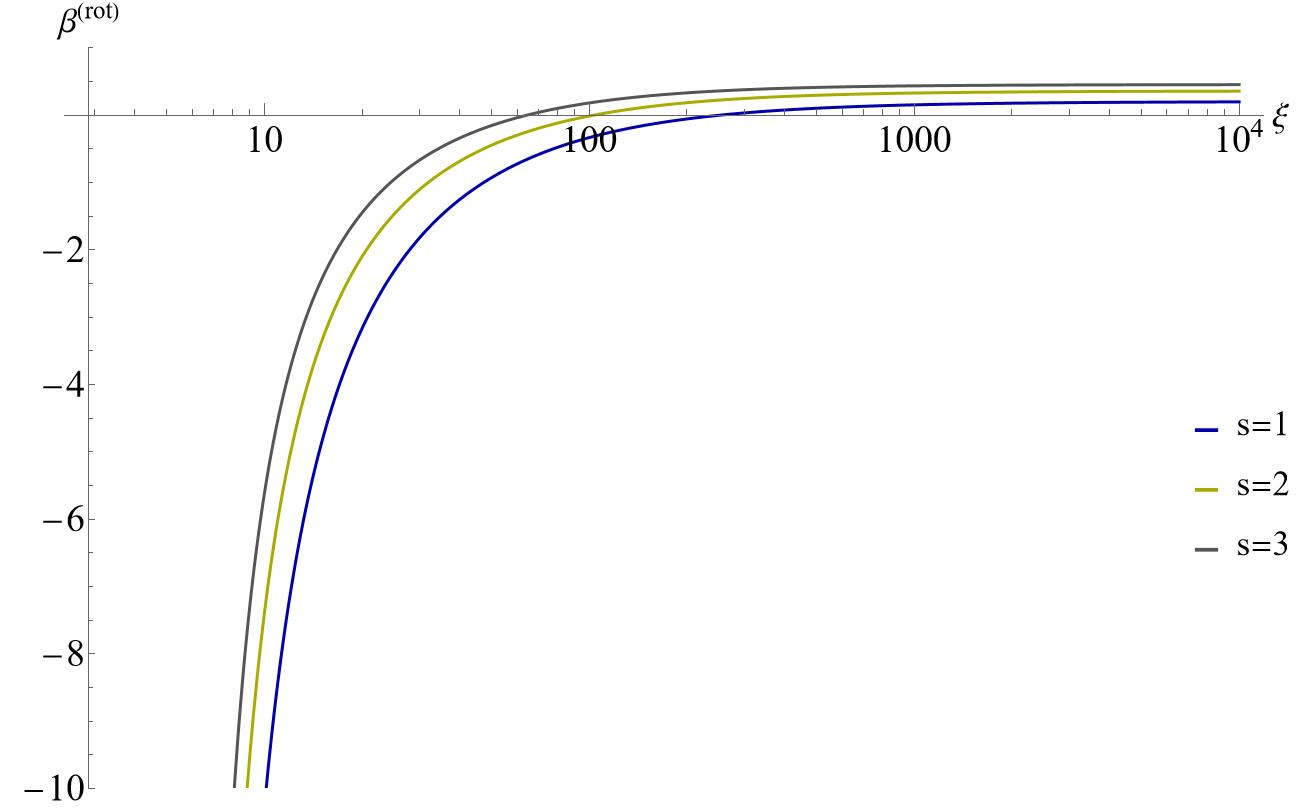}
    \end{subfigure}
    \caption{Log plot showing the behavior of the anisotropy parameter for the rotating model as a function of the dimensionless radius $\xi$ in the equatorial plane. Left panel: plot for the rotating gas model with parameter values $(s,\varepsilon_0)=(1,1)$ and different values of $k$. Right panel: plot for the rotating gas model with parameter values $(k,\varepsilon_0)=(6,1)$ and different values of $s$.}
    \label{Fig:Brot}
\end{figure}

Figure~\ref{Fig:Brot} shows the behavior of the anisotropy parameter as a function of the dimensionless radius $\xi$ in the equatorial plane for the rotating gas model (non-zero total angular momentum). In the left panel, it can be seen that the anisotropy parameter exhibits a clear dependence on the value of $k$, particularly in the intermediate $\xi$ region. For a fixed $\xi$, the anisotropy decreases as $k$ increases, indicating that, in the rotating case, the choice of the parameter $k$ significantly influences the macroscopic observables in the non-collisional regime. In the right panel, unlike the previous panel, the anisotropy parameter shows a moderate dependence on the value of $s$, especially for intermediate and large $\xi$. As $s$ increases, the anisotropy becomes slightly less negative, suggesting that the internal parameterization $s$ does have an effect in the rotating model, unlike what was observed in the non-rotating case.

In contrast to the non-rotating model, in the rotating configuration the anisotropy parameter is not exclusively negative. As shown in Figure~\ref{Fig:Brot}, for certain combinations of parameters and sufficiently large radii, the anisotropy parameter crosses zero and becomes positive. This indicates that the preferential direction in the gas velocity distribution can reverse in the presence of a total non-zero angular momentum, a feature that is absent in the non-rotating case. Furthermore, for large radii, the anisotropy parameter does not tend to zero but rather approaches a constant positive asymptotic value. This asymptotic limit depends on the parameter $s$ (which increases with $s$), while it remains independent of $k$.

A qualitative comparison between both models reveals several differences when contrasting the rotating case (Fig.~\ref{Fig:Brot}) with the non-rotating one (Fig.~\ref{Fig:Beven}). In the non-rotating model, the anisotropy was practically insensitive to variations in $s$ and asymptotically vanished at large radii, indicating full isotropization. In contrast, the rotating model exhibits a clear dependence on $s$, particularly in the intermediate and large $\xi$ regions, and approaches a positive constant value scaled by $s$. The dependence on $k$, on the other hand, is present in both models, though with subtle differences in magnitude. Most notably, the rotating model allows the anisotropy parameter to become positive for certain parameter combinations and maintains a finite asymptotic anisotropy, whereas the non-rotating model remains strictly negative throughout the entire domain and fully isotropizes at infinity. These observations suggest that the presence of total non-zero angular momentum introduces a richer parametric structure in the description of the anisotropy, affecting not only the sign and magnitude of the macroscopic observables but also their asymptotic behavior.

To further enrich the discussion, in the next section we will introduce another macroscopic observable of interest: the kinetic temperature. In contrast to the thermal temperature discussed in Section~\ref{Sec:Preliminaries}, this kinetic temperature can be constructed directly from the particle density and principal pressures presented above.

\section{Kinetic temperature behavior}
\label{Sec:KTemperature}

\begin{figure}[htbp]
    \centering
    \begin{subfigure}[b]{0.47\textwidth}
        \includegraphics[width=8cm]{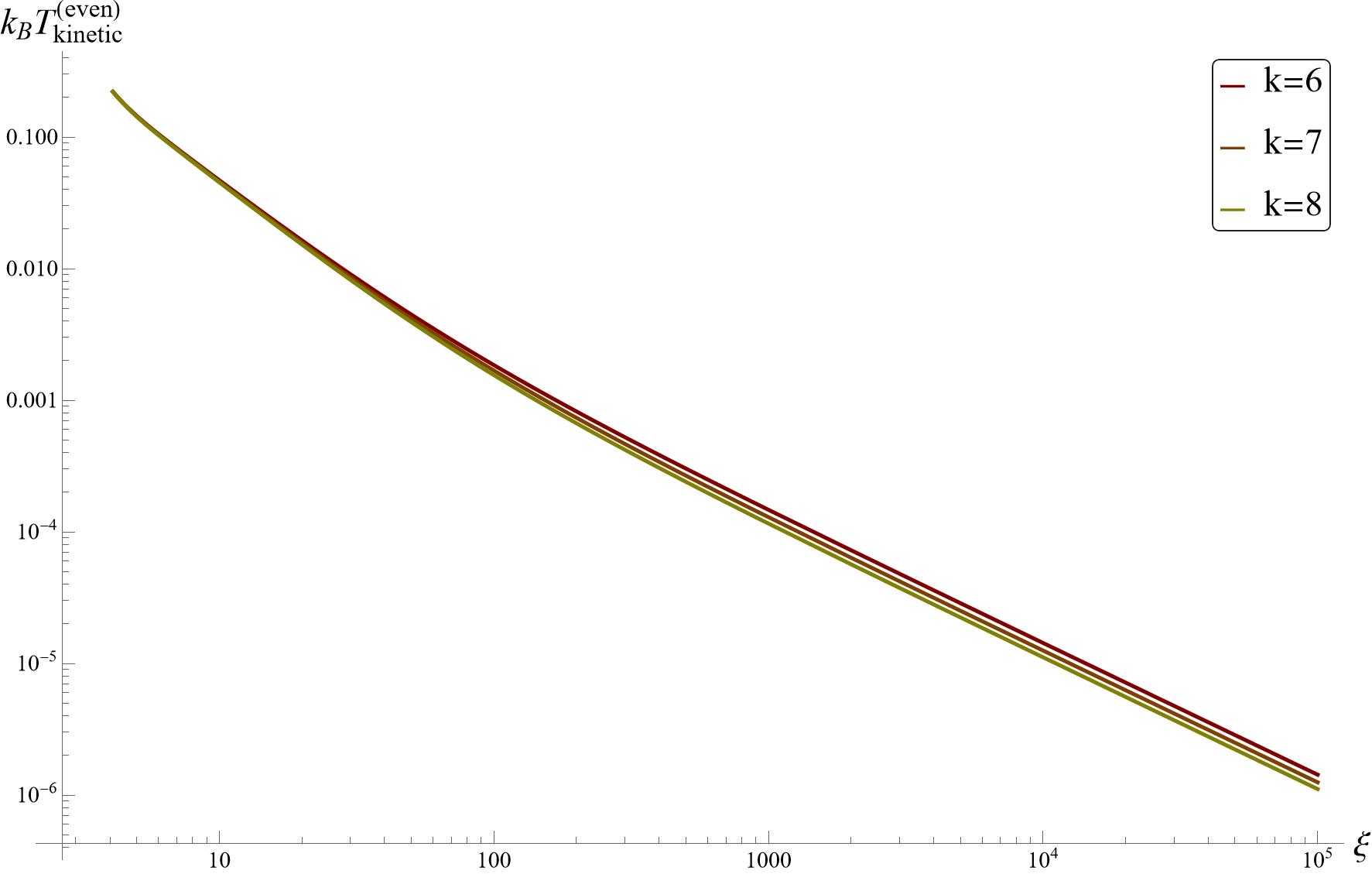}
    \end{subfigure}
    \hfill 
    \begin{subfigure}[b]{0.47\textwidth}
        \includegraphics[width=8cm]{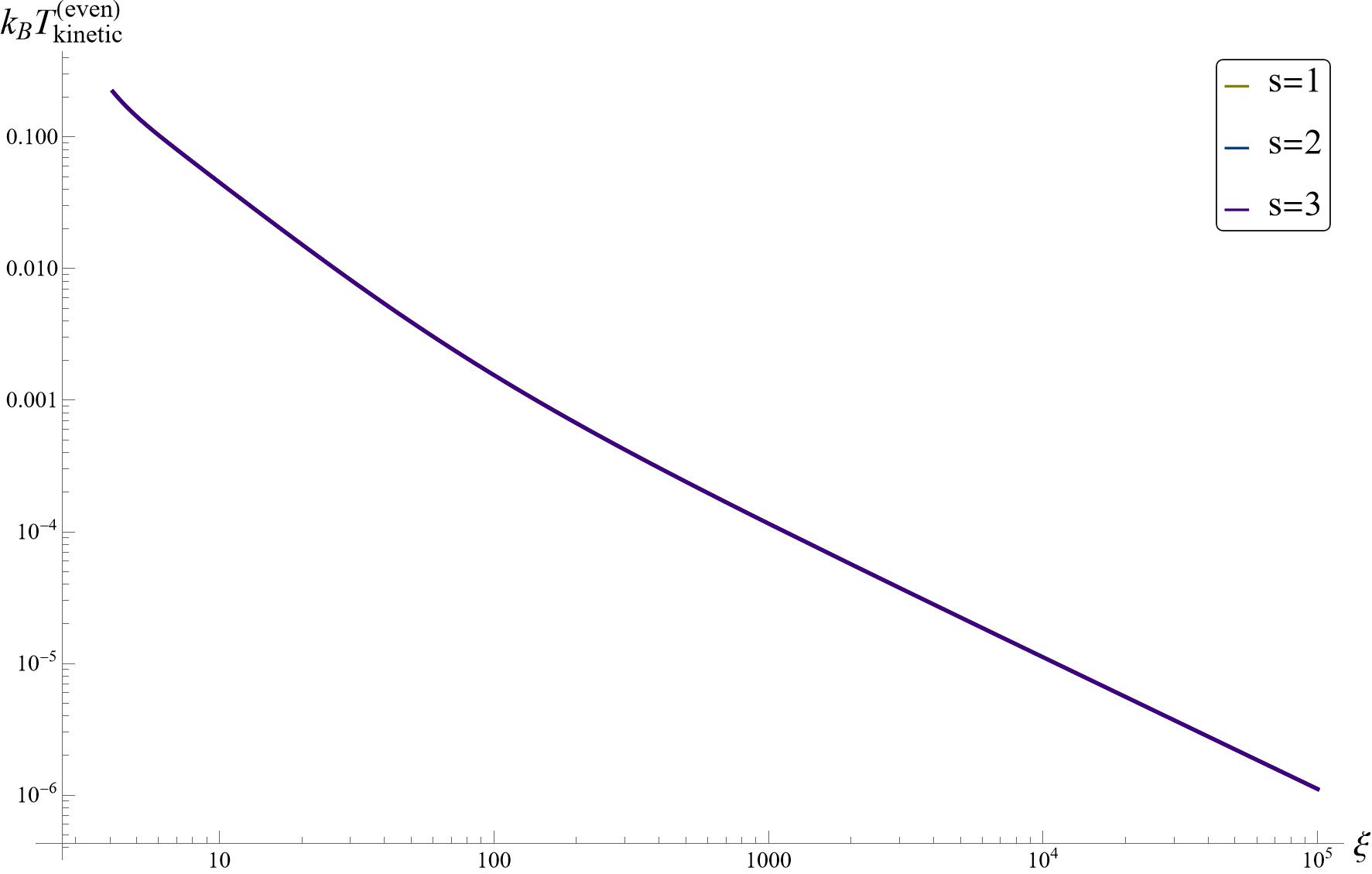}
    \end{subfigure}
    \caption{Log-log plot showing the behavior of the kinetic temperature for the non-rotating model as a function of the dimensionless radius $\xi$ in the equatorial plane. Left panel: plot for the non-rotating gas model with parameter values $(s,\varepsilon_0)=(1,1)$ and $k=6,7,8$. Right panel: plot for the non-rotating gas model with parameter values $(k,\varepsilon_0)=(6,1)$ and $s=1,2,3$.}
    \label{Fig:Teven}
\end{figure}

\begin{figure}[htbp]
    \centering
    \begin{subfigure}[b]{0.47\textwidth}
        \includegraphics[width=8cm]{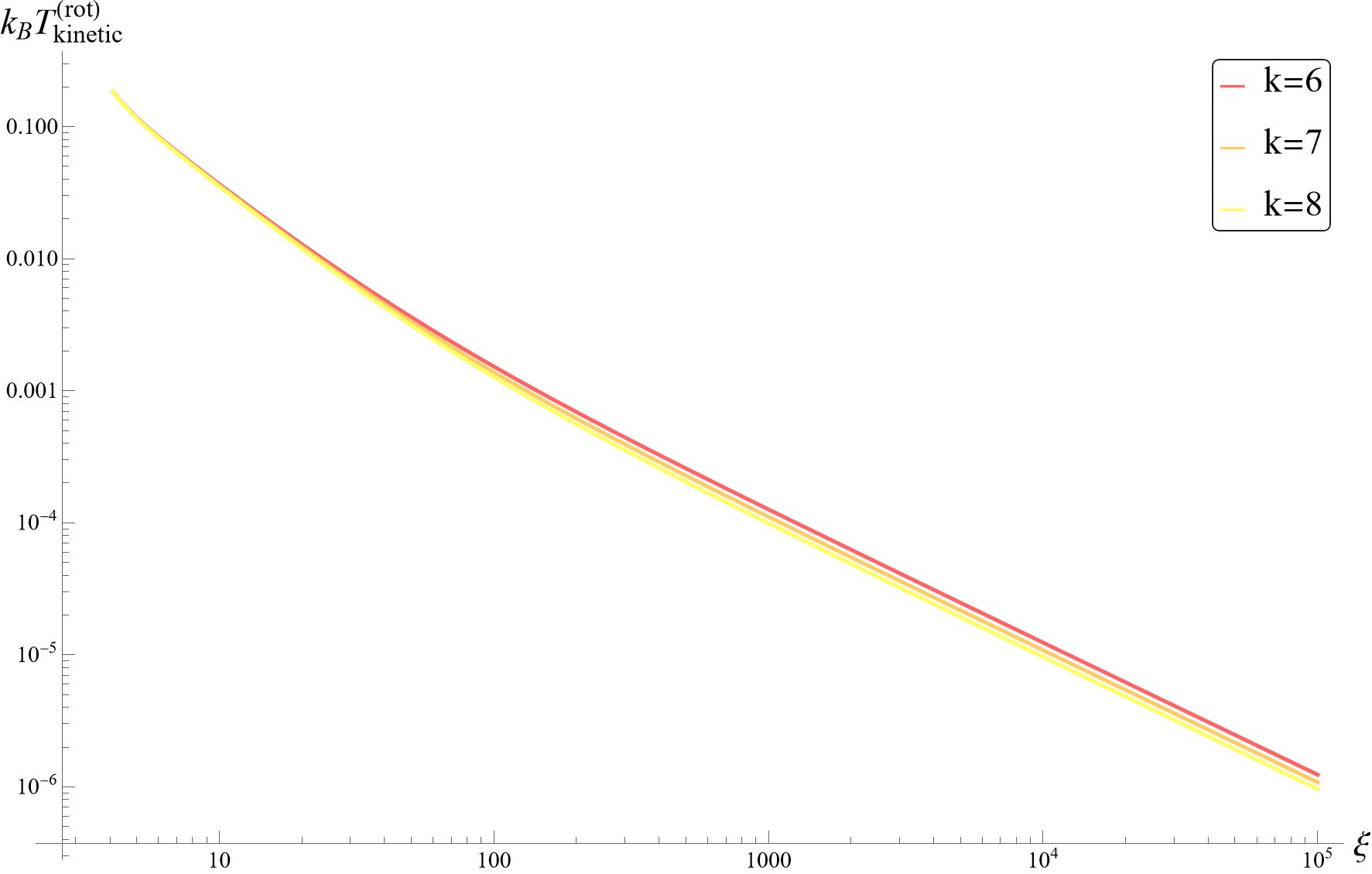}
    \end{subfigure}
    \hfill 
    \begin{subfigure}[b]{0.47\textwidth}
        \includegraphics[width=8cm]{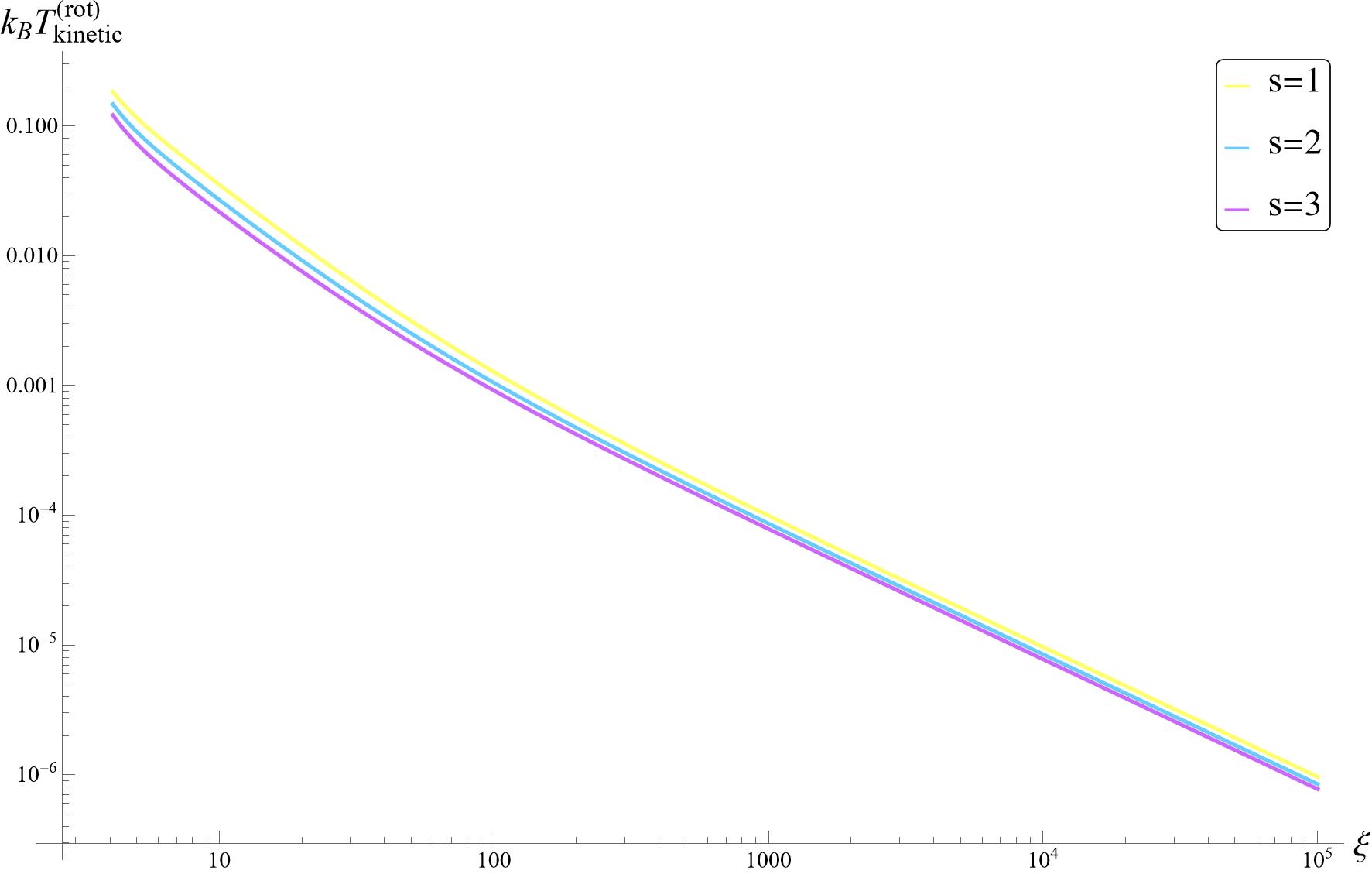}
    \end{subfigure}
    \caption{Log-log plot showing the behavior of the kinetic temperature for the rotating model as a function of the dimensionless radius $\xi$ in the equatorial plane. Left panel: plot for the rotating gas model with parameter values $(s,\varepsilon_0)=(1,1)$ and $k=6,7,8$. Right panel: plot for the rotating gas model with parameter values $(k,\varepsilon_0)=(6,1)$ and $s=1,2,3$.}
    \label{Fig:Trot}
\end{figure}

Since we are studying the properties and behavior of the macroscopic observables of a non-collisional gas, we cannot speak of or refer to a thermal temperature of the gas in any sense. Such a thermal temperature is established for gases in thermodynamic equilibrium, which are a response to the frequent collisions between the individual particles of the system~\cite{wI1967}. However, we are free to introduce a kinetic temperature through the ideal gas law, such that
\begin{equation}
    k_B T_{\textrm{kinetic}}^{(\textrm{even,rot})} = \frac{P_{\textrm{prom}}^{(\textrm{even,rot})}}{n^{(\textrm{even,rot})}}
\end{equation}
with $k_B$ being the Boltzmann constant, $n^{(\textrm{even,rot})}$ the particle density~(\ref{Eq:ParticleDensity}), and $P_{\textrm{kinetic}}^{(\textrm{even,rot})}$ the average pressure~(\ref{Eq:Pprom}). Similar analyses of this macroscopic observable can be found in~\cite{cGoS2023b} for the relativistic regime and in~\cite{cGoS2023a} for the non-relativistic case. Figures~\ref{Fig:Teven} show the behavior of the kinetic temperature as a function of the dimensionless radius $\xi$ for the non-rotating gas model, while Figures~\ref{Fig:Trot} correspond to the rotating case.

For the non-rotating model, left panel in figure~\ref{Fig:Teven} displays the kinetic temperature for fixed $(s,\varepsilon_0)=(1,1)$ and varying $k=6,7,8$. We observe a clear dependence on the parameter $k$: at a given radius $\xi$, the temperature decreases as $k$ increases. This sensitivity indicates that, even in the absence of angular momentum, the choice of $k$ significantly affects the macroscopic thermal state of the gas. In contrast, right panel in figure~\ref{Fig:Teven} shows the kinetic temperature for fixed $(k,\varepsilon_0)=(6,1)$ and varying $s=1,2,3$. Here, the curves are practically indistinguishable for all $\xi$, demonstrating that the temperature is insensitive to the internal parameter $s$ in the non-rotating configuration. This behavior mirrors that observed for the anisotropy parameter, reinforcing the idea that $s$ plays a minor role in the non-rotating regime.

For the rotating model, the situation is relatively different. The left panel in figure~\ref{Fig:Trot} presents the kinetic temperature for fixed $(s,\varepsilon_0)=(1,1)$ and varying $k=6,7,8$. In this case, the dependence on $k$ is weak; the curves almost overlap, with only marginal differences appearing at large $\xi$ (e.g., for $\xi=1000$, the value for $k=8$ is slightly lower than for $k=6$ and $k=7$). This suggests that in the presence of angular momentum, the kinetic temperature becomes nearly independent of $k$, unlike the non-rotating case. On the other hand, in the right panel, Figure~\ref{Fig:Trot} shows the temperature for fixed $(k,\varepsilon_0)=(6,1)$ and varying $s=1,2,3$. Here, the curves are again very similar, but a subtle dependence on $s$ can be detected at the largest radii (e.g., at $\xi=10000$ and $s=3$ yields a slightly higher value than $s=1$ or $s=2$). Thus, in the rotating model, the kinetic temperature exhibits a weak but non-zero sensitivity to both $k$ and $s$, with the dependence on $s$ becoming more noticeable in the asymptotic region. This contrasts with the non-rotating model, where only matter $k$, and highlights the richer parametric structure introduced by angular momentum.

Finally, a natural question is how these results would contrast with those obtained from a hydrodynamic description, where frequent collisions enforce local thermodynamic equilibrium. In such a regime, anisotropies are rapidly erased, pressures become isotropic, and the temperature follows a different radial profile, typically governed by an equation of state. Moreover, the effects of angular momentum in a collisional fluid are transmitted through collective motions rather than through the distribution function itself. Comparing both approaches highlights the distinctive signatures of collisionless dynamics: persistent anisotropies, sensitivity to the parameters, $k$ and $s$. The following discussion in section~\ref{Sec:HydroModel} outlines some qualitative points of contrast, helping to contextualize the kinetic results and to underscore the importance of the appropriate model choice depending on the physical regime of interest.

\section{Some comments in a comparison with the hydrodynamic model}
\label{Sec:HydroModel}

To characterize the macroscopic behavior of the kinetic model proposed in this article, we compare its predictions with those of a relativistic fluid model, such as the Polish doughnuts~\cite{Rezzolla-Book}. For this hydrodynamic model, one assumes an adiabatic fluid in local thermodynamic equilibrium that obeys the ideal gas equation $P = n k_B T$. The polytropic equation of state is given by $P = K n^\gamma$ where $K$ is a constant and $\gamma$ is the adiabatic index subject to $1 < \gamma \leq 2$. This adiabatic index $\gamma$ is related to the polytropic index $k$ by $\gamma = 1 + 1/k$ (see~\cite{BinneyTremaine-Book} and~(\ref{Eq:F0})). Here, the particle density $n$, temperature $T$, and pressure $P$ are related to the specific enthalpy $h$, (see for example Appendix E in~\cite{cGoS2023b} for more details) by
\begin{equation}
h - 1 
= \frac{\gamma}{\gamma-1}\frac{K}{\bar{m}} n^{\gamma-1}
= \frac{\gamma}{\gamma-1}\frac{k_B}{\bar{m}} T,
= \frac{\gamma}{\gamma-1} \frac{K^{1/\gamma}}{\bar{m}} P^{1-1/\gamma},
\end{equation}
where $\bar{m}$ is the averaged rest mass per particle. 

For a polytropic flow, the specific entropy per baryon remains constant throughout the fluid if the polytropic index $\gamma$ coincides with the gas adiabatic index, which is precisely our assumption. Using the first law of thermodynamics and the previous relations, one can derive an expression for the entropy. For a relativistic ideal gas, the specific entropy $s$ is related to $K$ and $\gamma$ by
\begin{equation}
s = \frac{1}{\gamma-1} \ln K + \text{constant}.
\label{Eq:SpecificEntropy}
\end{equation}
A Polish doughnut with a polytropic equation of state; the entropy is uniform throughout the torus (isoentropic flow). This implies $\nabla s = 0$, which is consistent with the barotropic fluid assumption. In its original and simplest formulation, the Polish doughnut model assumes that the disc matter can be described by the energy-momentum stress tensor of a perfect fluid. In such a fluid, the pressure is isotropic, meaning it is the same in all directions (radial, azimuthal, and vertical). Consequently, there is no intrinsic anisotropy parameter (as in the kinetic model), as the system is, by definition, isotropic in the fluid's rest frame. This assumption greatly simplifies the equilibrium equations (the relativistic Euler equation) and allows for the construction of analytical solutions for the shape and structure of the torus~\cite{Rezzolla-Book}.

Furthermore, to make or establish a qualitative comparison graphically, we adjust the adiabatic index $\gamma$ with the polytropic index $k$, so that the fluid and kinetic configurations are described with the same adiabatic index
\begin{equation}
    \gamma := \gamma_{\textrm{kinetic}} = 1 + \frac{1}{k}.
    \label{Eq:GammaKinetic}
\end{equation}

\begin{figure}[h!]
\centering
\includegraphics[width=0.5\linewidth]{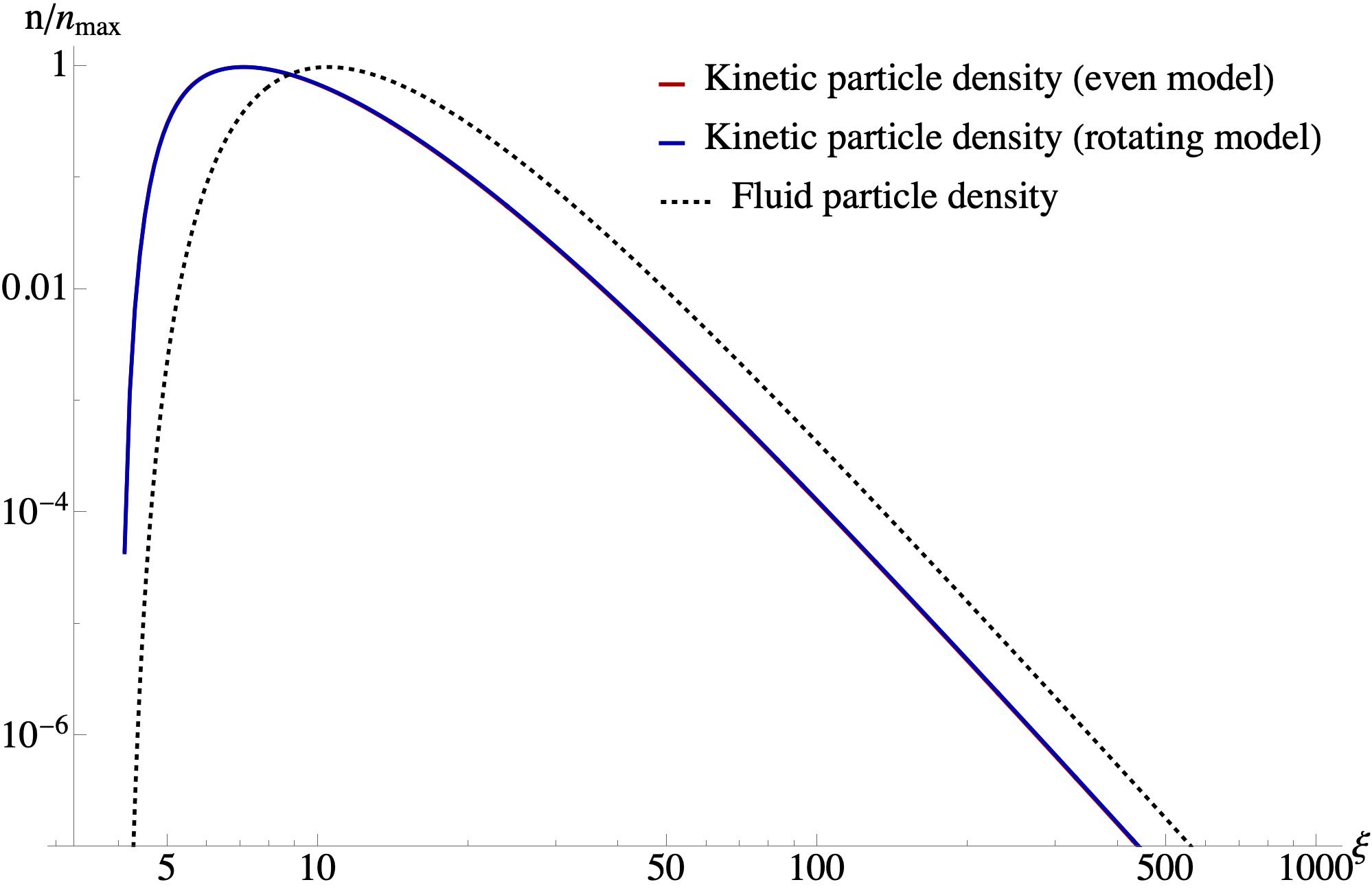}
\caption{Normalized profiles between the kinetic and fluid descriptions. The profile of the particle density in the fluid description is with an adiabatic index $\gamma=1.20$ (dotted line) and the corresponding particle density for the kinetic models (continuous lines). Both kinetic model profiles are plotted with $(k,s,\varepsilon_0)=(5,1,1)$.}
\label{Fig:NfluNkin}
\end{figure}

Figure~\ref{Fig:NfluNkin} shows the profile of the particle density for the kinetic and fluid models, both of them normalized over the maximum of their respective configurations. Both profiles are seen to share the same global qualitative shape: in each case the density increases from large radii up to a maximum and then decreases toward the inner region. This indicates that the fluid model and the kinetic gas configuration exhibit the same morphology. However, relevant systematic differences are present. In particular, the radial positions of the maxima do not coincide: the density maximum in the kinetic model is displaced regarding that of the fluid model, independent from the choice of parameters $(k, s)$ used in the kinetic gas model. The particle densities in the rotating and non-rotating kinetic models are superimposed, as they return the same behavior and there is no noticeable graphical difference.

\begin{figure}[h!]
\centering
\includegraphics[width=0.5\linewidth]{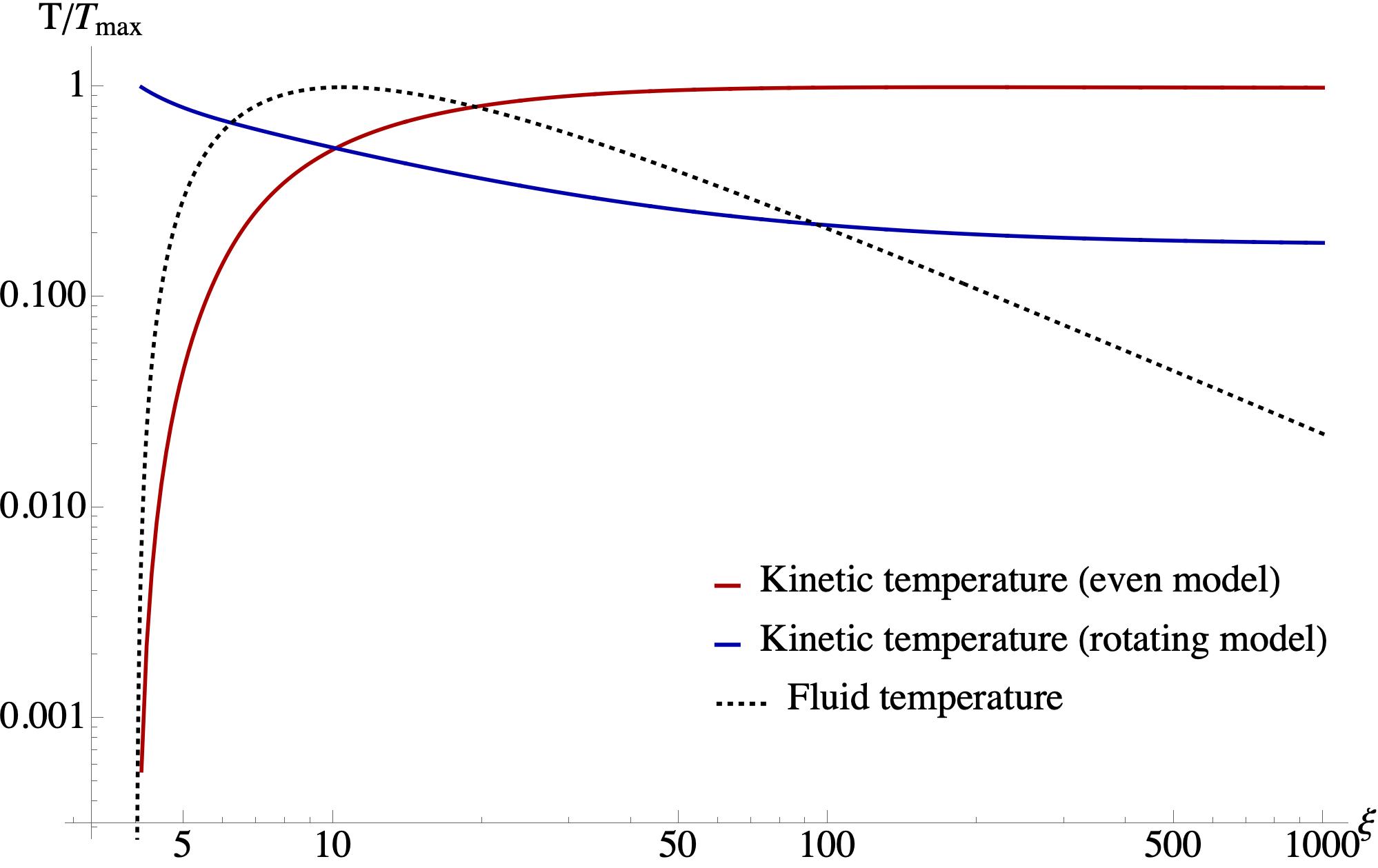}
\caption{Normalized profiles between the kinetic and fluid descriptions. The profile of the temperature in the fluid description is with an adiabatic index $\gamma=1.20$ (dotted line) and the corresponding kinetic temperature for the kinetic models (continuous lines). Both kinetic model profiles are plotted with $(k,s,\varepsilon_0)=(5,1,1)$.}
\label{Fig:TfluTkin}
\end{figure}

The comparison of the temperature profiles in Figure~\ref{Fig:TfluTkin} reveals a qualitatively different behavior. In contrast to the density case, no clear correlation is observed between the temperature distributions of the kinetic and the fluid models. The radial variations of one do not track those of the other, neither in the position of their maxima nor in the overall profile shape. In particular, the radii at which the temperature reaches characteristic values differ significantly between the two models, and regions where one temperature increases do not correspond to analogous regions in the other. Therefore, while in the density there exists at least a global morphological correspondence between both descriptions, in the temperature case the predictions of the fluid and kinetic models are different from each other. The same qualitative pattern is observed in rotating kinetic gas configurations constructed from non-even distribution functions. In such cases, although rotation modifies the radial location of the maxima and the width of the profiles, the comparison with the corresponding fluid model retains the same features: agreement in the morphology of the density of particle profiles but with displaced maxima and discrepancies at large radii, and absence of correlation between the fluid and kinetic temperatures for these models. This suggests that the differences observed between the fluid and kinetic descriptions do not depend on the absence of rotation but rather constitute a generic property of the collisionless kinetic model considered here.

\begin{figure}[h!]
\centering
\includegraphics[width=0.5\linewidth]{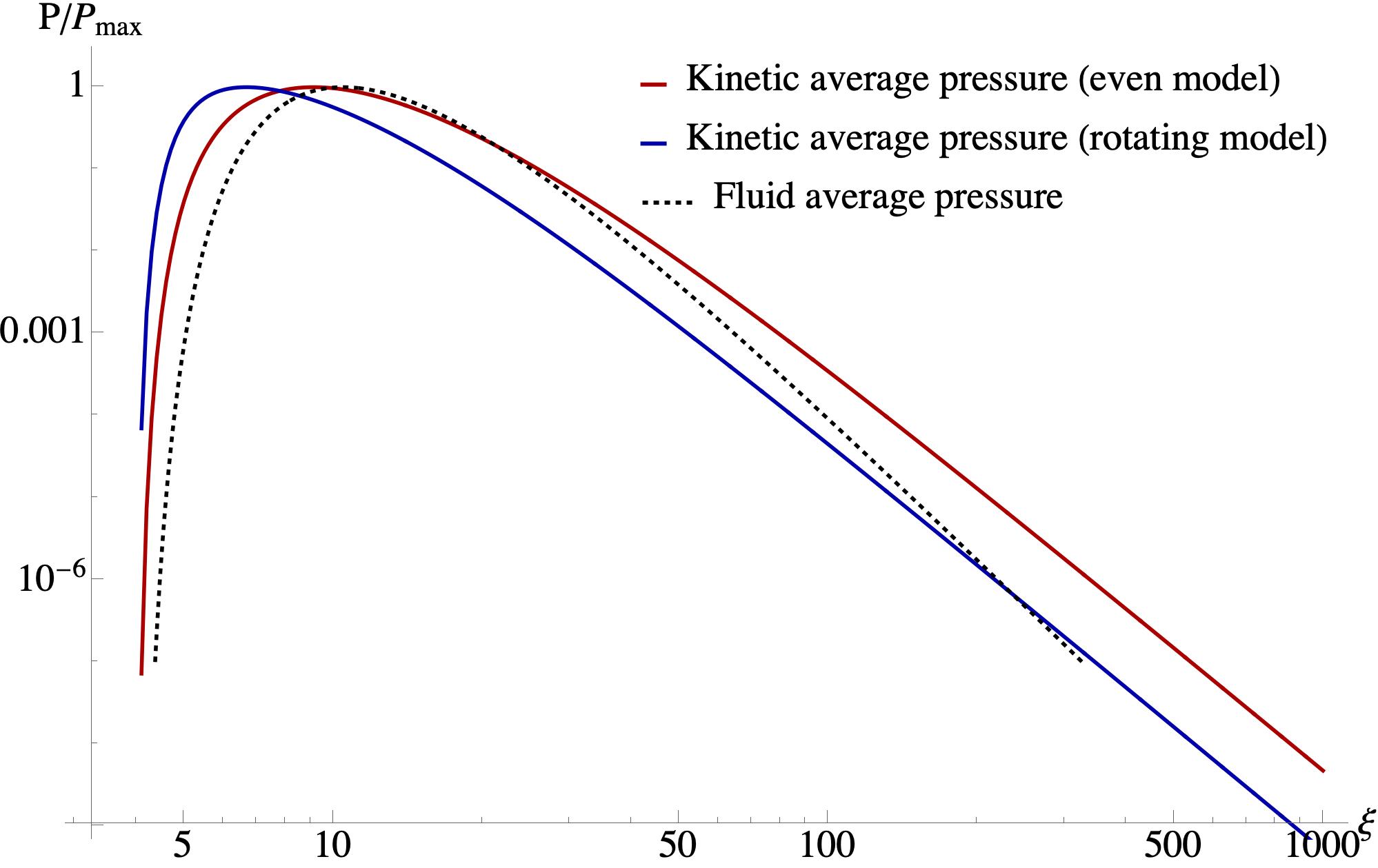}
\caption{Normalized profiles between the kinetic and fluid descriptions. The profile of the average pressure in the fluid description is with an adiabatic index $\gamma=1.20$ (dotted line) and the corresponding average pressure for the kinetic models (continuous lines). Both kinetic model profiles are plotted with $(k,s,\varepsilon_0)=(5,1,1)$.}
\label{Fig:PfluPkin}
\end{figure}

Figure~\ref{Fig:PfluPkin} shows the behavior of the average pressure as a function of the dimensionless radius $\xi$ for the three cases under consideration: the kinetic even (non-rotating) model, the kinetic rotating model, and the fluid (hydrodynamic) model. All three curves exhibit virtually identical behavior across the entire radial domain. For small radii the pressure increases, while in the intermediate region it is reaching a maximum. Beyond large radii, the pressure decreases sharply, approaching zero. The striking agreement between the kinetic models and the fluid description indicates that, despite the fundamentally different underlying dynamics between collisionless versus collisional systems, the average pressure is an observable that remains insensitive to both the presence of angular momentum and the detailed kinetic parameters. This suggests that the average pressure is primarily determined by the overall density distribution and gravitational potential, rather than by the specific microscopic state of the gas.

Based on the results shown so far, the next chapter will state the conclusions derived from this study.

\section{Conclusions}
\label{Sec:Conclusions}

In this article, we have studied the morphological properties of a collisionless relativistic kinetic gas around a Schwarzschild black hole, described by a one-particle distribution function that depends on the constants of motion through an ansatz motivated by the inclination angle of bound orbits and a generalized polytropic energy dependence~\cite{cGoS2022,cGrR2025}. By considering two distinct models for the angular dependence, the even model describing a non-rotating gas and the rot model describing a rotating one, we have analyzed and compared the behavior of key macroscopic observables, including the entropy density, the anisotropy parameter, the kinetic temperature, and the average pressure.

We have derived explicit expressions for the components of the entropy covector field for both models, including the angular integrals provided in Appendix~\ref{Appx:A} that determine their dependence on the polar angle. The normalized invariant entropy density, defined in Eq.~(\ref{Eq:BarS}) and shown in Figs.~\ref{Fig:Seven} and~\ref{Fig:Srot}, reveals that configurations with larger values of the polytropic parameter $k$ exhibit higher entropy density levels, while increasing the angular parameter $s$ concentrates the entropy more sharply. A novel result is the entropic contrast between models presented in Fig.~\ref{Fig:SevenSrot}: the ratio $S^{(\mathrm{even})}/S^{(\mathrm{rot})}$ is strictly greater than unity throughout the entire radial domain, demonstrating that the presence of angular momentum systematically reduces the invariant entropy density. This ratio exhibits a minimum at intermediate radii ($\xi \sim 10^2$) and increases monotonically toward large distances, indicating that the entropy suppression induced by rotation persists even in the asymptotic region and is not merely a strong-field effect.

The anisotropy parameter, defined from the principal pressures, shows markedly different behavior between the two models, as illustrated in Figs.~\ref{Fig:Beven} and~\ref{Fig:Brot}. In the non-rotating case, it is always negative, insensitive to variations in $s$, and tends asymptotically to zero at large radii, indicating complete isotropization far from the black hole. In contrast, the rotating model exhibits a clear dependence on both parameters: it decreases with increasing $k$ and shows a moderate sensitivity to $s$, particularly at intermediate and large $\xi$. Most notably, the anisotropy parameter in the rotating model can cross zero and become positive for certain parameter combinations, and it approaches a constant positive asymptotic value that depends exclusively on $s$ while remaining independent of $k$. This constitutes a key finding: rotation introduces a persistent residual anisotropy that does not vanish even at large distances, fundamentally distinguishing rotating from non-rotating collisionless gases.

The kinetic temperature, constructed from the ideal gas law using the average pressure and the particle density, exhibits contrasting parametric sensitivities, as shown in Figs.~\ref{Fig:Teven} and~\ref{Fig:Trot}. In the non-rotating model, it depends significantly on $k$ but is practically insensitive to $s$, mirroring the behavior of the anisotropy parameter. In the rotating model, however, the temperature shows weak sensitivity to both parameters, with a subtle dependence on $s$ becoming noticeable only at the largest radii. This indicates that angular momentum affects the anisotropy and modulates the thermal response of the gas, smoothing out the parametric dependence observed in the non-rotating case.

Finally, we have performed a systematic qualitative comparison between our kinetic models and a relativistic fluid description, specifically the Polish doughnut model~\cite{Rezzolla-Book} with an adiabatic index $\gamma = 1 + 1/k$ matched to the kinetic polytropic parameter via Eq.~(\ref{Eq:GammaKinetic}). The results, summarized in Figs.~\ref{Fig:NfluNkin}-\ref{Fig:PfluPkin}, reveal a panorama of its comparison. The particle density profiles show global morphological agreement between kinetic and fluid models, both increasing from large radii to a maximum and then decreasing inward; however, the radial position of the maximum is systematically shifted between the two descriptions, a discrepancy that persists regardless of the choice of parameters $k$ and $s$. The kinetic temperature displays no correlation with the fluid temperature: the radial variations, maxima locations, and overall profile shapes are qualitatively different. Remarkably, the average pressure exhibits a similar behavior across all models (kinetic non-rotating, kinetic rotating, and fluid) throughout the entire radial domain.

In summary, this work provides a comprehensive characterization of the macroscopic observables of collisionless relativistic kinetic gases around Schwarzschild black holes, revealing how the inclusion of angular momentum qualitatively modifies the entropy, anisotropy, and temperature profiles. The anisotropy in the rotating model, the construction of the flux entropy covector, and the contrasting behavior of the kinetic temperature compared to fluid descriptions constitute original contributions that extend previous studies. These findings highlight the importance of a kinetic treatment for collisionless systems in strong gravitational fields and open the door to further investigations, such as the inclusion of weak collisions, self-gravity effects, or extensions to other spacetime backgrounds for the bound orbits.

\acknowledgments

We thank Olivier Sarbach for fruitful comments and discussions throughout this work. C. Gabarrete acknowledges support from SECIHTI through grant CBF-2025-G-1626.

\appendix
 
\section{Performing angular integrals in kinetic models}
\label{Appx:A}

The $\mathcal{I}$ functions that appear in the components of the entropy flux covector field, which determines the angular dependency of the macroscopic observables in the non-rotating or rotating models explored in this article, are defined as follows: 
\begin{eqnarray}
    \label{Eq:Iieven}
    \mathcal{I}^{(\textrm{even})}_{i}(\vartheta) &:=& \int\limits_0^{2\pi} G_{i}^{(\textrm{even})}(\vartheta,\chi) d\chi 
    = \int\limits_0^{2\pi} (\sin\vartheta \sin\chi)^{2s} d\chi \nonumber \\
    &=& 2\sqrt{\pi}\frac{\Gamma(s+1/2)}{\Gamma(s+1)}\sin^{2s}\vartheta \\ 
    \label{Eq:Ii2rot}
    \mathcal{I}^{(\textrm{rot})}_{i/2}(\vartheta) &:=& \int\limits_0^{2\pi} G^{(\textrm{rot})}_{i/2}(\vartheta,\chi) d\chi 
    = \frac{1+s}{1+2s}\frac{1}{2^s}\int\limits_0^{2\pi} (1 + \sin\vartheta \sin\chi)^{s} d\chi \nonumber \\
    &=& \frac{2\pi}{2^s} \frac{1+s}{1+2s} {}_{2}F{}_{1}\left(\frac{1-s}{2} , -\frac{s}{2} ; 1 ; \sin^2\vartheta \right) \\
    \label{Eq:Itildeieven}
    \widetilde{\mathcal{I}}_i^{(\textrm{even})}(\vartheta) &:=& \int\limits_0^{2\pi} G_i^{(\textrm{even})}(\vartheta,\chi) \log\left[ G_i^{(\textrm{even})}(\vartheta,\chi)\right] d\chi 
    = \int\limits_0^{2\pi} (\sin\vartheta \sin\chi)^{2s}\log\left[(\sin\vartheta \sin\chi)^{2s}\right] d\chi \nonumber \\
    &=& 2 s \sqrt{\pi} \frac{\Gamma(s+1/2)}{\Gamma(s+1)} \left[ \Psi_{\textrm{di}}\left(s+\frac{1}{2} \right) - \Psi_{\textrm{di}}\left(s+1 \right) + 2\log\sin\vartheta \right] \sin^{2s}\vartheta \\
    \label{Eq:Itildei2rot}
    \widetilde{\mathcal{I}}_{i/2}^{(\textrm{rot})}(\vartheta) &:=& \int\limits_0^{2\pi} G_{i/2}^{(\textrm{rot})}(\vartheta,\chi) \log\left[G_{i/2}^{(\textrm{rot})}(\vartheta,\chi)\right] d\chi 
    = \frac{1+s}{1+2s}\frac{1}{2^s} \int\limits_0^{2\pi} \left(1+\sin\vartheta \sin\chi \right)^{s}\log\left[\frac{1+s}{1+2s}\frac{1}{2^s} \left(1+\sin\vartheta \sin\chi \right)^{s}\right] d\chi \nonumber\\
    &=& \frac{2\pi}{2^s}\frac{1+s}{1+2s} \left[ \log\left(\frac{1+s}{1+2s}\frac{1}{2^s}\right) + s \frac{\partial}{\partial s} \right] {}_2F_1 \left(\frac{1-s}{2},-\frac{s}{2};1;\sin^2\theta\right) \\
    \label{Eq:Ihati2rot}
    \widehat{\mathcal{I}}^{(\textrm{rot})}_{i/2}(\vartheta) &:=& \int\limits_0^{2\pi} G_{i/2}^{(\textrm{rot})}(\vartheta,\chi) \sin\chi d\chi 
    = \frac{1+s}{1+2s} \frac{1}{2^s} \int\limits_0^{2\pi} (1 + \sin\vartheta\sin\chi)^{s} \sin\chi d\chi \nonumber \\
    &=& \frac{\pi s}{2^s}\frac{1+s}{1+2s} {}_{2}F{}_{1}\left(\frac{1-s}{2} , 1-\frac{s}{2} ; 2 ; \sin^2\vartheta \right) \sin\vartheta \\
    \label{Eq:Ibari2rot}
    \overline{\mathcal{I}}_{i/2}^{(\textrm{rot})}(\vartheta) &:=& \int\limits_0^{2\pi} G_{i/2}^{(\textrm{rot})}(\vartheta,\chi) \log\left[G_{i/2}^{(\textrm{rot})}(\vartheta,\chi)\right] \sin\chi  d\chi 
    = \frac{1+s}{1+2s}\frac{1}{2^s} \int\limits_0^{2\pi} \left(1+\sin\vartheta \sin\chi \right)^{s}\log\left[\frac{1+s}{1+2s}\frac{1}{2^s}\left(1+\sin\vartheta \sin\chi \right)^{s}\right] \sin\chi d\chi \nonumber\\
    &=& \frac{\pi s \sin\vartheta}{2^s}\frac{1+s}{1+2s} \left[ 1 + \log\left(\frac{1+s}{1+2s}\frac{1}{2^s}\right) + s \frac{\partial}{\partial s} \right] {}_2F_1\!\left(\frac{1-s}{2},\frac{2-s}{2};2;\sin^2\vartheta\right)
\end{eqnarray}
Here ${}_{p}F{}_{q}(a_1, \ldots, a_p ; b_1, \ldots, b_q ; z)$ is the generalized hypergeometric function and $\Psi_{\textrm{di}}$ denotes the digamma function (see~\cite{DLMF-Book}) defined by
\begin{equation}
\Psi_{\textrm{di}}(k) := \frac{\partial}{\partial k} \ln \Gamma(k), \qquad k>0. 
\end{equation}

The analytical expressions for the angle integrals in Eqs.~(\ref{Eq:Itildei2rot}) and~(\ref{Eq:Ibari2rot}) were initially proposed using artificial intelligence tools, as they could not be evaluated via conventional analytical methods. These expressions were subsequently verified through numerical integration in Wolfram Mathematica~\cite{Mathematica}, showing convergence to three significant figures. For the purposes of this work, this level of precision is considered sufficient. The authors confirm the validity of the results presented and assume full responsibility for the work.


\bibliography{ref.bib} 

\end{document}